\documentclass[12pt,preprint]{aastex}

\usepackage{psfig}

\begin{document}
\def\newpage{\vfill\eject}
\def\min{{\rm min}}
\def\max{{\rm max}}
\def\kpc{{\rm kpc}}
\def\pc{{\rm pc}}
\def\rpm{{\rm RPM}}
\def\kms{{\rm km}\,{\rm s}^{-1}}
\def\masyr{{\rm mas}\,{\rm yr}^{-1}}
\title{Disk and halo wide binaries from the Revised Luyten Catalog:
\\probes of star formation and MACHO dark matter}

\shorttitle{Revised Luyten Wide Binaries}

\author{Julio Chanam\'e\, and Andrew Gould} 
\affil{{}Department of Astronomy, The Ohio State University,
Columbus, OH 43210, USA}
\email{jchaname,gould@astronomy.ohio-state.edu}

\begin{abstract}

We present a catalog of 1147 candidate common proper motion binaries
selected from the revised New Luyten Two-Tenths Catalog (NLTT).  Among
these, we identify 999 genuine physical pairs using the measured
proper-motion difference and the relative positions of each binary's
components on a reduced proper-motion (RPM) diagram.  The RPM
positions also serve to classify them as either disk main-sequence
(801), halo subdwarf (116), or pairs containing at least one white
dwarf (82).  The disk and halo samples are complete to separations of
$\Delta\theta=500''$ and $\Delta\theta=900''$, which correspond to
$\sim 0.1\,$pc and $\sim 1\,$pc, respectively.  At wide separations,
both distributions are well described by single power laws, $d
N/d\Delta\theta \propto (\Delta\theta)^{-\alpha}$: $\alpha=1.67\pm 0.07$
for the disk and $\alpha=1.55\pm 0.10$ for the halo.  The fact that
these distributions have similar slopes (and similar normalizations as
well) argues for similarity of the star-formation conditions of these
two populations.  The fact that the halo binaries obey a single power
law out to $\sim 1\,$pc permits strong constraints on halo dark-matter
candidates.  At somewhat closer separations ($10''\la \Delta\theta \la
25''$), the disk distribution shows a pronounced flattening, which is
detected at very high statistical significance and is not due to any
obvious systematic effect.  We also present a list of 11 previously
unknown halo stars with parallaxes that are recognized here as
companions of Hipparcos stars.

\end{abstract}


\setcounter{footnote}{0}
\renewcommand{\thefootnote}{\arabic{footnote}}

\section{Introduction
\label{sec:intro}}

Binaries form under the influence of their original star-cluster
environment, including both stars and gas, but after the dissolution
of their parent cluster, they remain mostly undisturbed during their
subsequent several-Gyr voyages through the Galaxy.

There are a few exceptions to this rule.  If one or both members
evolve off the main sequence (MS), then the accompanying mass loss (or
mass transfer) can influence the orbit.  Very close binaries (a
$\lesssim 0.1$ AU) can circularize due to tidal interaction.  Very
wide binaries (a $\gtrsim 100$ AU) are so weakly bound that they can
be significantly disturbed, even disrupted, by the extremely weak
perturbations from inhomogeneities in the Galactic potential due to
stars, molecular clouds, dark objects, or large-scale tides.  Wide
binaries of unevolved stars are clearly subject to only the last of
these three effects.

A carefully chosen sample of unevolved wide binaries can therefore
shed light both on their process of formation and the graininess of
the Galactic potential through which they travel.  To be useful, the
sample need not be complete, it need only be free of strong selection
effects as a function of projected separation.  It would be of
particular interest to obtain substantial samples of disk and halo
binaries using the {\it same} selection procedure.  These two
populations may have formed under very different conditions and they
have subsequently probed very different parts of the Galactic
potential.  Hence, comparison of the binary distribution functions of
disk and halo samples chosen by the same procedure could help throw
important light on both populations.

To date, the wide-binary distribution function has been measured by
two closely related methods: two-point correlation and common proper
motion (CPM).  The first approach relies entirely on photometric data.
One first measures the overall density of stars within a catalog of
known completeness properties and then measures the excess of near
neighbors as a function of angular separation relative to the
background expected from random unassociated ``optical pairs''.  Since
the density of optical pairs per unit separation $\Delta\theta$ grows
$\propto\Delta\theta$, this method would appear to fail as soon as the
number of real pairs in a $\Delta\theta$ bin falls below the
square-root of the number of the optical pairs.  In fact, the method
can be pushed slightly farther by using two-band photometric data: one
can require that the two components have the same photometric distance
based on the assumption that both are on the MS.  Of course, this
eliminates real pairs with one component that is white dwarf (WD) or a
giant star, but it reduces the background by a factor of a few.  The
same photometric distances can then be used to determine the binary's
projected physical separation, $r_\perp$, from its angular separation,
$\Delta\theta$.  \citet{bs81} pioneered this method, obtaining a list
of 19 candidate pairs from a sample of 3000 stars assembled by
\citet{weistrop}.  \citet{gould95} applied this technique to a sample
drawn from {\it Hubble Space Telescope} data to obtain 13 candidate
pairs, and based on this made the first estimate of the relative rate
of disk and halo wide binaries from a single homogeneous sample.
\citet{garn88,garn91} applied this technique to a much larger sample,
obtaining puzzling results that we discuss later in \S 4.3.  See also
the Appendix of \citet{gould95}.  Once candidates are found using the
two-point correlation technique, they can be confirmed by radial
velocity (RV) measurements, as was done by \citet{latham} for some of
the stars in the \citet{bs81} sample.

The second (CPM) method adds proper-motion information to the
positional astrometry and photometry of the first.  Because of their
wide separations and corresponding low orbital velocities, wide
binaries should have very similar proper motions and, if these are
available, they can be used to distinguish genuine binaries from much
more numerous pairs of unassociated stars in the field.  Of course,
the technique does break down at sufficiently large separations,
primarily because the number of field stars eventually becomes so
large that some unassociated stars actually have similar proper
motions, but ultimately because at sufficiently large separations, the
observed proper motions represent substantially different projections
of the common physical velocity.  \citet{ww} applied this technique to
the Woolly catalog and, after vetting their sample using RV
measurements, \citet{crc} obtained a sample of 32 wide binaries.

Multiplicity studies as a function of stellar type in the
Galactic-field population include the works of \citet{duq91} for G
dwarfs and \citet{fis92} for M dwarfs.  These two important works
combine observational techniques that operate on different regimes of
angular separation to assemble volume-limited samples of binaries
spanning more than 10 decades of orbital period.  While both studies
find similar period distributions for these spectral types (showing a
Gaussian-type shape as a function of $\log P$ with a broad peak
centered around $P\sim40$ yr, or equivalently $a\sim25$ AU), the
M-dwarf binary frequency ($\sim 42\%$) is lower than that of G dwarfs
($\sim 57\%$), although this could be due to the somewhat different
ranges of companion masses for the two samples.  Pre-MS stars,
however, seem to differ significantly.  Several surveys of T Tauri
stars ($\lesssim 5$ Myr) in nearby star-forming regions like
Taurus-Auriga, Ophiucus-Scorpius, Chameleon, Lupus, and Corona
Australis, find a binary fraction about twice as large as that of
field stars (\citealt{ghez93}; \citealt{leinert93}; \citealt{simon95};
\citealt{ghez97}; \citealt{white01}).  There is, nevertheless, one
interesting exception, the Orion Nebula cluster (ONC), for which
\citet{petr98} find a binary fraction in agreement with that of MS
field stars (see also \citealt{prosser94}; \citealt{scally99}; and
\citealt{duchene99}).  Since the ONC is different than the other
surveyed star-forming regions in that it has a much higher stellar
density, this result suggests an environmental dependence of the
binary fraction among pre-MS stars.

Recent works have focused on the binary properties of MS populations
as a function of age and environment, and their comparison to what is
observed in the field.  While the comparison between T Tauri and
Galactic field populations reveals a discrepancy between the measured
binary fractions of pre-MS and old MS stars, multiplicity studies in
stellar clusters with young MS populations such as $\alpha$ Persei
($\sim 90$ Myr), the Pleiades ($\sim 120$ Myr), and the Hyades and
Praesepe ($\sim 660$ Myr) find a binary fraction consistent with that
of the corresponding range of separations among the older G and M
dwarfs in the field.  Thus, the binary frequency appears not to
decline with age on timescales from $\sim 90$ Myr to $\sim$ 5 Gyr
(\citealt{bouvier97}; \citealt{patience98}; \citealt{patience02}).

None of the above works, however, has attempted the construction and
subsequent comparison of complete samples of binaries unambiguously
belonging to the disk and halo of the Galaxy.  The difficulty in
assembling statistically significant samples of binaries with
well-understood selection effects, together with the requirement of
obtaining the information necessary to classify the stars as belonging
to either population, must have certainly worked against this goal.
Nevertheless, a few important efforts to obtain samples of either disk
or halo binaries should be mentioned.  First, in the most recent
report of a large spectroscopic survey of high proper-motion stars,
\citet{latham02} find no difference in the binary frequency and period
distribution of two samples of (93) disk and (78) halo binaries.
These are, however, close binaries with periods of less than 20 yr
(their spectroscopic monitoring started in 1987) or, equivalently,
semi-major axes $a \lesssim$ 7 AU.  \citet{ryan92} obtained optical
photometry of a number of CPM pairs selected from the New Luyten
Two-Tenths (NLTT) Catalog, which he determined to belong to the
Galactic halo.  Since a fraction of his binaries are common to the
present work, we leave the discussion of this paper to \S 5.3.  Also,
Allen, Poveda, \& Herrera (2000) searched in NLTT for CPM companions
of a sample of high-velocity and metal-poor stars taken from another
source, finding 122 binaries.  Here we only note that this corresponds
to just $\sim 5\%$ of all the wide binaries available in NLTT, and
discuss the \citet{allen} work in detail in \S 5.3.

From this summary (see also Appendix of \citealt{gould95}), it appears
that obtaining a large sample of gravitationally bound wide binaries
is a formidable task.  Generally surveys are sensitive to only about 2
decades of separation, being bounded by merging images at the close
end and by confusion with unassociated field stars at the wide end.
The total fraction of stars with wide companions is only a few percent
per decade of separation \citep{duq91}, and within a given survey many
of these may be lost due to the magnitude limits of the survey.

The NLTT Catalog \citep{luy,1st} would seem to be an obvious source of
wide binaries.  NLTT is a proper-motion limited catalog
($\mu>180\,\masyr$), which is largely complete over 19 magnitudes.  It
contains photographic photometry in two bands for the great majority
of its almost 59,000 entries.  Most importantly in the present
context, it is appended by a set of notes which, in particular,
identify pairs believed by Luyten to be CPM binaries\footnote{Another
Luyten catalog, the LDS (Luyten Double Star), contains all the 6121
candidate double stars with common proper motion discovered by Luyten
over a half century, including those below the NLTT proper motion
threshold of $\mu>180\,\masyr$.}.  It would probably have been
possible to carry out a complete search for wide binaries in NLTT, and
it is not completely clear to us why no one has attempted to do this,
since the sample obtained would be several orders of magnitude larger
than those previously listed.  However, we do note several obstacles
that would have had to have been overcome.  First, the NLTT
photographic colors are quite crude and so do not permit the
construction of a reliable reduced proper motion (RPM) diagram
\citep{rpm}.  Hence, the NLTT photometry is only marginally useful in
discriminating genuine binaries from random field pairs.  Second, for
most cases for which Luyten believed that a pair was a CPM binary, he
did not record his independent proper-motion measurements.  Instead he
assigned both stars the same proper motion.  As we show below, a
significant fraction of the widest binaries ($\Delta\theta \gtrsim
100\arcsec$) so recorded are not genuine pairs, but rather are
unassociated stars with substantially different proper motions (see \S
2.4 and Fig.1).  However, because Luyten did not record his
measurements, there is no way to identify these except by remeasuring
their proper motions.  Third, Luyten failed to identify a number of
genuine binaries at wide separations because his measurement errors of
$\sigma\sim 20\,\masyr$ \citep{faint} did not permit him to reliably
distinguish these from the numerous unrelated optical pairs at these
separations.  Thus, considerable additional work would have been
required to extend an NLTT-based sample into the range
$\Delta\theta\ga 100''$.

With the publication of the revised NLTT (rNLTT) by \citet{bright} and
\citet{faint}, all three of these problems can be overcome.  For the
44\% of the sky covered by the intersection of the first Palomar
Observatory Sky Survey (POSS I) and the Second Incremental Release of
the Two-Micron All-Sky Survey (2MASS \citealt{2mass}), rNLTT has
optical-infrared photometry for the great majority of NLTT stars, and
this is sufficient to permit good stellar classification using an RPM
diagram \citep{rpm}.  The roughly 3-fold increase in temporal baseline
of rNLTT relative to NLTT permits a substantial improvement in
proper-motion accuracy to $\sigma\sim 5.5\,\masyr$.  Moreover, the
accuracy of {\it relative} proper motions of nearby stars (which is
what is relevant to the problem of CPM binaries) is even better:
$\sigma \sim 3\,\masyr$.  This means that the vector proper motion
measurements of real pairs typically differ by only $6\,\masyr$
\citep{faint}.  Finally, rNLTT contains entries comparing Luyten's
original (circa 1950) estimate of the vector separation of the binary
components with rNLTT's own (circa 2000) measurement, and these can be
compared to help identify spurious CPM binaries.

Nevertheless, it is by no means straightforward to assemble a
uniformly selected catalog of wide binaries from the rNLTT.  First,
for binaries with separations $\Delta\theta\la 10''$, rNLTT often
fails to identify both companions because in its underlying sources,
mostly USNO-A \citep{usnoa1,usnoa2} and 2MASS, these can be blended at
such close separations.  Even at wider separations, NLTT stars can be
missing from one or both of USNO-A and 2MASS for a variety of reasons,
including crowding, saturation, as well as other often unidentifiable
effects.  If it is missing from both, then the NLTT star is not in
rNLTT.  If it is missing from one, then rNLTT does not have an
independent proper motion.  Finally, the Second Incremental 2MASS
release has almost fractal sky coverage.  Hence, one component of a
binary might be in rNLTT and the other not simply because they lie on
opposite sides of this very complicated boundary.  Since the
probability of this happening increases with separation, it must be
carefully investigated.

Here we assemble a catalog of wide rNLTT binaries.  In proper motion
surveys, binaries are identified by means of the CPMs of their
components, which in rNLTT are measured over a timescale of about 45
years (the time between the acquisition of the POSS-I plates and the
2MASS survey).  These binaries have not appreciably changed their
apparent separation on the sky in half a century, so their orbital
periods are almost all longer than several hundred years,
corresponding to semi-major axes above about 100 AU.

We take a variety of steps to systematically minimize all the
potential problems described above.  We classify binaries as either
disk or halo, and extend our search to separations
$\Delta\theta=500\arcsec$ for disk binaries and
$\Delta\theta=900\arcsec$ for halo binaries\footnote{We note at this
point that binaries with WD components are not considered in the final
samples from which our conclusions are derived.  This is because WDs,
being intrinsically less luminous, can be detected only over a limited
volume, one that is comparable to the volume over which MS stars can
be detected but much smaller than that probed by SDs.}.  These
correspond to physical separations of about 0.1 pc and 1 pc
respectively.  We find that both distributions of angular separations
are characterized by power laws $f(\Delta\theta)\equiv
dN/d\Delta\theta \propto (\Delta\theta)^{-\alpha}$, with
$\alpha=1.67\pm0.07$ and $\alpha=1.55\pm 0.10$ respectively.  That is,
the slopes are consistent with each other at the $1\sigma$ level.  We
also find that the normalizations are reasonably consistent.  The two
populations of wide binaries are also similar with respect to their
distributions of luminosity and mass ratios.  {\it Our main
conclusions are therefore that disk and halo binaries probably formed
under similar conditions, despite the very different metallicities and
epochs of formation.}

Furthermore, we find no evidence of any falloff in the number of
binaries at the largest separations with respect to these power laws.
Since the widest binaries are easy to disrupt in encounters with other
objects, our results allow for the possibility of establishing limits
on the density of massive dark objects in the Galactic halo.  Such
limits are derived in a companion paper \citep{jerry}.

We note here that, although the binaries studied in this work are all
of wide physical separations (a $\gtrsim 100$ AU), throughout this
paper and for reasons of convenience that will become clear in \S 2
and \S 3, we subdivide our dataset into ``close'' and ``wide''
subsamples, according to whether their angular separations on the sky
are smaller or larger than $10\arcsec$.

The structure of the paper is as follows.  In \S 2 we describe the
limits of our search for CPM systems and the selection of the
candidate pairs.  These are then classified in \S 3 as either genuine
disk or halo binaries.  The characteristics of the final clean samples
are studied in \S 4, where we also derive the distributions of angular
and physical separation, as well as the distributions of luminosity
and mass ratio.  In \S 5 we discuss our findings, and we present our
conclusions in \S 6.  The catalog description is given in Appendix A,
and in Appendix B we present a list of 12 previously unknown halo
stars with parallaxes that are recognized in this work as probable
companions of Hipparcos stars.  The catalog in ASCII format can be
retrieved from
http://www.astronomy.ohio-state.edu/$\sim$gould/rNLTT$\_$binaries/binaries.dat.gz.

\section{Sample Selection
\label{sec:select}}

\subsection{Overview
\label{sec:overview}}

Our overall strategic aim is to assemble a catalog of all genuine wide
binaries that satisfy three conditions.  First, both components are
catalogued in NLTT.  Second, at least one component is in the
intersection of POSS I and the 2MASS Second Incremental Release, i.e.,
the area over which rNLTT attempted to be complete\footnote{At the
time when the bulk of the work presented in this paper was finished,
the 2MASS All Sky Release was not yet available, but it was released
when preparing the manuscript.  For reasons of consistency, all the
results presented in this work make use of the Second Incremental
Release only.}.  Third, the separation is restricted to the range
$\Delta\theta<500''$ for disk binaries and $\Delta\theta<900''$ for
halo binaries.  We also exclude all triples.  Of course, one would
like to incorporate high proper-motion binaries with one or both
components missing from NLTT, but there are at present no publicly
available data from which these could be identified with a reasonable
amount of work.

To achieve this aim, we first investigate all pairs identified as
binaries in the NLTT notes (hereafter ``Luyten binaries'') to
determine if they are genuine and then check all other pairs up to the
maximum separation to determine whether any of these are also genuine
binaries.  At first sight, it may appear that the division into Luyten
and non-Luyten binaries is superfluous.  However, as we will soon
show, there is a wealth of information on the Luyten binaries that is
not available for the non-Luyten binaries, so we are really quite
fortunate that Luyten was very systematic in identifying binaries.

We then divide the 1073 Luyten binaries into two broad classes,
``close'' and ``wide'' according to whether Luyten measured their
separation as $\Delta\theta\leq 10''$ or $\Delta\theta>10''$.  These
contain respectively 37\% and 63\% of the sample.  This division is
motivated by two considerations.  First, the close binaries can be
regarded as real because the chance of two unassociated stars lying so
close in both position and velocity space is small: only of order one
such spurious binary is expected in the entire rNLTT catalog.
(Indeed, we searched for pairs within $10''$ that Luyten did not call
``binaries'' and found only one (NLTT 39521/39525).  We confirmed that
this one was indeed an unassociated pair.)  Second, the rate of
non-identifications in USNO-A and 2MASS is much higher among the close
than the wide binaries, so this is a natural division from the
standpoint of classifying binaries in the sample.  That is, for the
close binaries, we simply assume that they are all genuine, and focus
our efforts on classification (\S~\ref{sec:closeclass}), while for the
wide binaries, we assemble all the available information to first
determine if they are good CPM candidates (\S~\ref{sec:widegen}) and
then to classify them (\S~\ref{sec:wideclass}).

\subsection{Wide Luyten Binaries, $\Delta\theta>10''$
\label{sec:widegen}}

\begin{figure}
\vspace*{0cm}
\hspace*{-18cm}
\plotfiddle{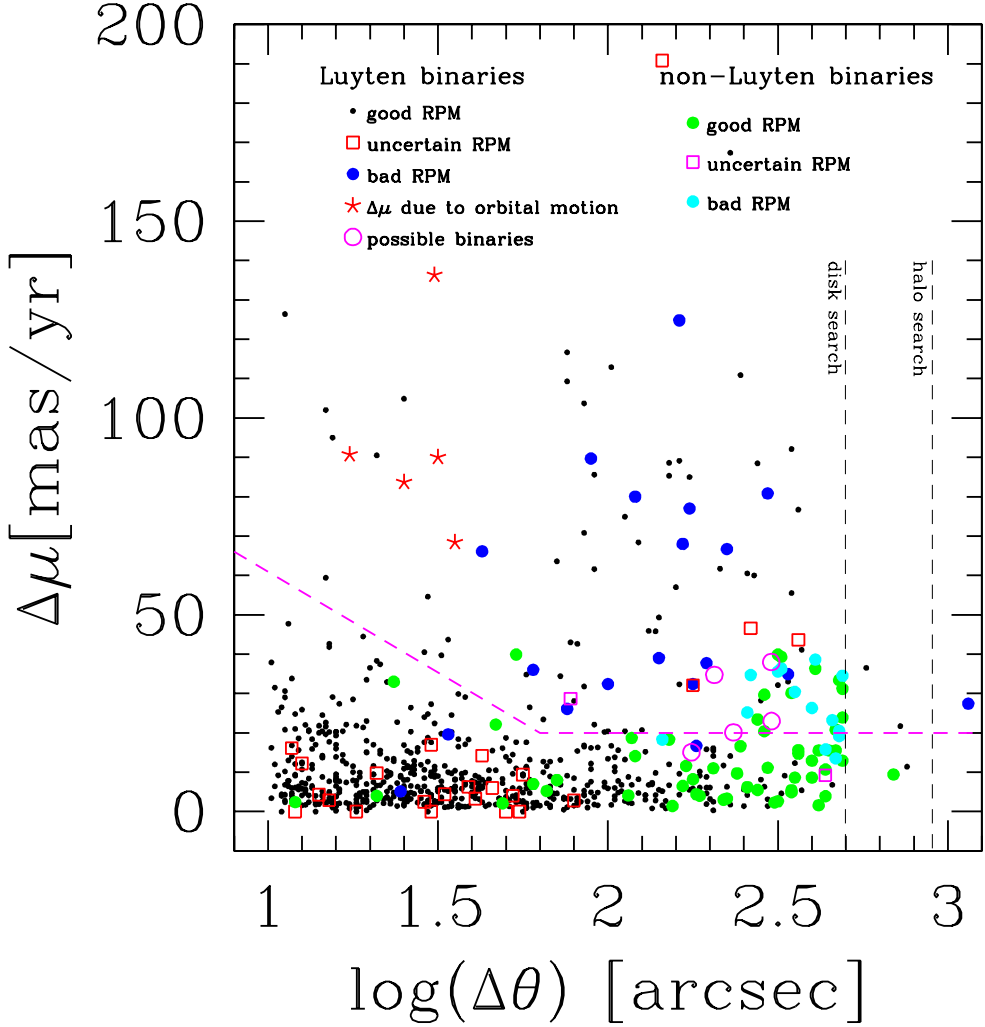}{1cm}{0}{0.7}{0.7}{0}{0}
\caption{\label{fig:one} Vector proper-motion difference versus
angular separation for the CPM systems with $\Delta\theta >
10\arcsec$.  The dashed magenta line delineates our selection criteria
for this ``wide subsample'': CPM pairs below this limit are
automatically accepted as candidate binaries for their subsequent
classification.  CPM pairs with proper-motion differences above the
cutoff but inside $50\arcsec$ were individually investigated.  Red
asterisks indicate those whose large proper-motion difference is
explained by significant orbital motions, and are accepted as good
candidates for classification.  The various colors and symbols encode
the binaries as being genuine, of uncertain type, or bad (i.e.,
unrelated pairs), for both Luyten and non-Luyten classes, after they
were subject to the classification procedure.  Magenta circles
indicate pairs that Luyten believed ``might be'' binaries.  The dashed
vertical lines indicate the upper limits of the search for disk and
halo binaries.  }\end{figure}

We begin the vetting of Luyten wide binaries by plotting the absolute
value of the vector proper-motion difference of the components against
separation.  We adopt the notation $\Delta{\rm\bf\mu} =
|\Delta{\rm\mbox{\boldmath$\mu$}}|$.  See Figure \ref{fig:one}.  It
goes without saying that binaries can only be plotted on this diagram
if both components have proper motions independently measured by
rNLTT.  However, this simple criterion fails for 138 of the 680 wide
Luyten binaries because one or both components have only the Luyten
proper-motion measurements available.  This can happen for any of the
following reasons: neither component in rNLTT (2), one component not
in rNLTT (43), one component lacking a proper motion because it is
identified only in 2MASS (82), one component lacking a proper motion
because it is identified only in USNO-A (11).

Fortunately, it is still possible to obtain independent proper-motion
measurements for the largest subcategory, the 82 stars with 2MASS-only
identifications in rNLTT.  These stars were originally identified in
rNLTT by searching in 2MASS for the companion to an already identified
rNLTT star at the position offset given by the NLTT notes.
\citet{faint} showed that, for angular separations smaller than
$57\arcsec$, these position offsets are generally accurate to $\sim
600\,\rm mas$, although there are some outliers (see their fig.\ 8).
Hence, by comparing the vector difference in epoch 2000 positions from
rNLTT with the vector offset given in the NLTT notes, we obtain an
estimate for the proper-motion difference with an error of $600\,\rm
mas/45\,\rm yr \sim 13\,\masyr$.  Although for pairs with separations
larger than $57\arcsec$, Luyten's original truncation of the position
angle to integer degree can lead to a substantially larger error, we
find below that this has negligible impact for the pairs relevant to
this discussion.  Moreover, by performing the same type of search
among the 37 binaries with only one component in rNLTT, we find an
additional 2 2MASS-only companions that were evidently missed by
\citet{rpm}.

We select stars as good candidates for further investigation according
to whether they fall above or below the dashed line in
Figure~\ref{fig:one}.  At wide separations, this curve is set at
$\Delta\mu=20\,\masyr$, corresponding to the $3\,\sigma$ error in
proper-motion differences (except for the 2MASS-only stars).  At
closer separations, we are more tolerant of large proper-motion
differences, partly because the chance of contamination by
unassociated pairs is lower (we quantify this issue in \S 2.4) and
partly because some pairs at close separations can have significant
orbital motions.  In fact, by individually exploring all the pairs in
Figure 1 that show proper-motion differences larger than the imposed
cut, but have separations smaller than $50\arcsec$, we identify 5
cases for which the large $\Delta\mu$ is due to orbital motions, and
these are indicated as red asterisks.  We further discuss this
selection in \S~\ref{sec:nonluy}, below.

We search for each of the remaining $43-2=41$ unidentified companions
to rNLTT stars in USNO-B \citep{usnob}.  \citet{gould03} showed that
this catalog is about 90\% complete for high proper-motion stars and
that its internally reported errors are generally accurate.  He also
showed that the catalog cannot be used to search in the field for high
proper-motion stars because about 99\% of its high proper-motion
entries are spurious.  However, it can be used to look for individual
high proper-motion stars at known locations.  Of these 41 companions,
we find 23 in USNO-B, and exclude 4 of these on the grounds of large
proper-motion difference between the components.  Finally, among those
not found in USNO-B, we exclude the 6 widest pairs ($\Delta\theta >
200\arcsec$) since no proper-motion difference between their
components can be established.

The remaining 12 binaries whose secondaries could not be identified
either in 2MASS or USNO-B need to be handled carefully, since the
primary may have been falsely identified by rNLTT.  We deal with these
in \S~\ref{sec:wideclass} where we include classification criteria to
assess their reality.  Here we just remark that 4 from this group make
it into our final clean sample.

Among the group of 82 binaries with 2MASS-only components discussed
above, we look in USNO-B for those wider than $25\arcsec$ and for
which the $\Delta\mu$ estimate from the vector positions at the two
epochs falls outside our selection cut (i.e., above the dashed line in
Fig. 1).  We find that USNO-B confirms large proper-motion differences
in all but 4 cases.  We mark these 4 pairs for further examination in
\S 3.2.

We search in USNO-B for the 11 wide pairs with a component lacking a
proper motion because it is only identified in USNO-A but not in
2MASS, and we find 8.  Seven of these pairs are confirmed to be CPM
systems, while the other has a large proper-motion difference and is
excluded from the sample.  Of the 3 pairs not found in USNO-B, one has
a separation of $11\arcsec$ so the NLTT proper motion errors
($\sigma_{\mu}=20\,\masyr$) are adequate given our $50\,\masyr$
tolerance at this separation (see Fig. 1), another has a very blue
component according to Luyten photometry so we designate it as a WD,
and the third is rejected because of insufficient information.

Finally, 8 binaries lack 2MASS photometry for both components, even
though there are entries for them in this survey.  The 16 stars in
these pairs are very bright Tycho objects that saturated the 2MASS
detectors.  With 2 exceptions, we confirm the reality of these pairs
as CPM systems by comparing images at two different epochs (ESO Online
Digitized Sky Survey, http://archive.eso.org/dss/dss).  In
\S~\ref{sec:wideclass} we will assign these six binaries to the disk
population, because of the brightness of both their components.

Note that the exclusion from the analysis of pairs outside the allowed
region of the $\log\Delta\theta - \Delta\mu$ plane does not mean that
all these systems should be regarded as false binaries.  Rather, it
reflects our philosophy of being as complete and free of selection
effects as possible, but at the same time rigorous in not being
contaminated by unrelated optical pairs.  CPM systems excluded by this
cutoff, especially those at large apparent separations, are good
candidates for radial velocity follow-up in order to determine their
reality.

\subsection{Non-Luyten Binaries
\label{sec:nonluy}}

Figure 1 also shows all pairs of rNLTT stars that satisfy the
following three conditions.  First, they are not listed as binaries in
the NLTT notes.  Second, they have rNLTT proper-motion differences
$\Delta\mu<40\,\masyr$.  Third, they have separations
$\Delta\theta<500''$.  Among halo stars, we also show those with
separations up to 900$\arcsec$ and $\Delta\mu<20\,\masyr$.  The
density of unassociated pairs scales $d^2 N/d\Delta\theta\,d\Delta\mu
\propto \Delta\theta\,\Delta\mu$ (see \S 2.4).  Hence, one expects the
unassociated pairs to be concentrated near $\Delta\mu \sim 40\,\masyr$
and $\Delta\theta\sim 500''$, while the real binaries should hover
about the x-axis.  When the Luyten and non-Luyten pairs are combined,
this expected behavior is evident.  It is clear that Luyten did indeed
miss some binaries, which we extract by imposing the same selection
indicated by the dashed curve.  It is also clear from the figure that
we cannot push our sample too far beyond $\Delta\theta=500''$ without
risking serious contamination.  We will discuss this point further in
\S~\ref{sec:wideclass}.

Note that in order for non-Luyten binaries to make it into our sample,
both components must have independent rNLTT proper-motion
measurements.  For comparison, of the 680 wide Luyten binaries, 138
had a least one companion initially lacking such a proper-motion
measurement.  However, for 87\% of these, we were able to {\it use the
information from the NLTT notes} that a companion existed as well as
its approximate offset, to actually locate the companion.  That is, we
have high-precision relative proper-motion measurements for 97\% of
the 680 wide Luyten binaries.  However, if NLTT stars have other NLTT
stars as genuine binary companions, but these are not recovered by
rNLTT, we have no way to identify them by searching in rNLTT alone.
Fortunately, we can use USNO-B to search for this last subset of
non-Luyten binaries.  The results of this search are discussed in \S
3.2.

\subsection{Number of false pairs at large angular separations
\label{false}}

With the search for CPM companions reaching increasingly larger
angular separations, the chance of contamination by unassociated pairs
that happen to show similar proper motions also increases.  This is
why we impose a rigorous upper limit in proper-motion difference at
large angular separations (Figure 1).

In order to quantify the number of false pairs as a function of
angular separation we perform a search in rNLTT for CPM pairs out to
angular separations of $2000\arcsec$ and proper-motion differences
$\Delta\mu < 60\,\masyr$.  Plots of the cumulative distributions of
the number of pairs ($N$) as a function of $\Delta\theta$ and
$\Delta\mu$ show that $N$ roughly scales as the square of both these
variables, as is expected (for a uniform density of pairs) just from
the increase in the four-dimensional phase-space volume.  Based on the
821 pairs found with separations between $1000\arcsec$ and
$2000\arcsec$ and $\Delta\mu < 60\,\masyr$, we infer a density of
unrelated optical pairs of

\begin{equation}
\frac{d^{4}N}{d\Delta\theta_x\,d\Delta\theta_y\,d\Delta\mu_x\,d\Delta\mu_y}
\simeq 7.7\times10^{-9}\,{\rm arcsec^{-2}}\,(\masyr)^{-2}\,.
\end{equation}

For our allowed region between $100\arcsec$ and $500\arcsec$ and
$\Delta\mu < 20\,\masyr$ (see Fig. 1), this density implies that about
7 false pairs should survive our initial selection criteria.  Indeed,
as can be verified in Figure 1 with the help of the color codings,
there are 6 pairs inside this region that we classify (\S 3) as either
bad or uncertain matches, and which are not included in the final
samples.  This confirms the reliability of the combination of our
selection and classification criteria in sorting out the physical
binaries from the large background of false unassociated pairs.

In contrast, for the same interval of angular separations but with
$20\,\masyr < \Delta\mu < 40\,\masyr$ (i.e., outside our selection
box), this same density implies that $\sim 22$ of the total of 37
pairs in this region are due to chance alignments.

Note that, because halo stars comprise only $\sim 1/4$ of rNLTT, the
density of false halo pairs is a factor of 16 times lower than that
given by equation (1).  Hence, within our $\Delta\theta < 900\arcsec$
limit, we expect about one false halo pair, before vetting by relative
position in the RPM diagram (see \S 3 and \S 5.1).

\section{Classification: Disk Versus Halo Binaries
\label{sec:classify}}

The large size of the binary sample selected in \S~\ref{sec:select} is
by itself a significant improvement over previous efforts.  However,
our main motivation is to exploit the unprecedented capability of
rNLTT to cleanly separate local disk and halo populations.  These two
advantages combine to give us, for the first time, complete and
unbiased samples of wide binaries belonging to stellar populations
whose formation and evolution remain as open questions in Galactic
astronomy.  In this section we describe our classification
methodology. We also show how this procedure permits the
identification of the few unrelated pairs and bad matches that slipped
through the selection criteria of \S~\ref{sec:select}.

We separate the disk and halo populations using the RPM discriminator
$\eta$ introduced by \citet{faint},

\begin{equation}
\eta = V_{\rm RPM} - 3.1(V-J) - 1.47\,|\sin\,b\,| - 2.73,
\label{eqn:etadef}
\end{equation}

\noindent where

\begin{equation}
V_{\rm RPM} = V + 5\,\log\mu,
\label{eqn:rpmdef}
\end{equation}

\noindent is the RPM\footnote{This expression for the RPM is different
than that used by Luyten, who would write $H_{\rm pg} = m_{\rm pg} +
5\,\log\mu + 5$, where the subscript stands for ``photographic''.  We
choose here to follow the notation used in the original rNLTT papers
\citep{bright,faint}.}, $\mu$ is the proper motion of the star in
arcsec per year, $b$ is its Galactic latitude, and $V-J$ and $V$ are
its color and magnitude.  Basically, we classify stars as disk or halo
according to whether $\eta$ is negative or positive.

To understand how the RPM discriminator $\eta$ works, we examine each
term separately.  The RPM is $V_\rpm=M_V+ 5\log(v_\perp/47.4\,\kms)$,
where $v_\perp$ is the transverse speed. Hence, the RPM is a proxy for
the star's intrinsic luminosity: if all stars had the same speed, the
RPM would equal the absolute magnitude $M_V$ plus an additive
constant.  Since halo stars typically move about 5 times faster than
disk stars relative to the Sun and since they are typically a
magnitude or two fainter at fixed color, their RPMs are typically
about 5 magnitudes greater than those of disk stars.  Hence, despite
the considerable dispersions in $V_\rpm$ for each population, the halo
and disk tracks are well separated on a RPM diagram (see \citealt{rpm}
and Figure 2 in the present paper).

The second term, $3.1(V-J)$ is approximately the slope of the ``blank
track'' between the two tracks populated by the disk MS and the halo
SDs.  That is, the equation of this blank track is $V_\rpm = 3.1(V-J)
+ {\rm const}$.  However, as demonstrated by figure 2 of
\citet{faint}, this constant is actually a function of Galactic
latitude $b$.  At higher latitude, the transverse motions of both halo
and disk stars are larger than at low $b$.  This is partly because of
their asymmetric drift in the direction of Galactic rotation and
partly because of their higher dispersion in the radial direction.
The third term takes account of this effect.  The last term is
included for convenience, that is, so that $\eta=0$ at the boundary.

In fact, $\eta$ discriminates not only between MS stars and SDs but
also WDs: stars are classified as disk if $\eta < 0$, as halo if $0 <
\eta < 5.15$, and as WDs if $\eta > 5.15$.

\begin{figure}
\vspace*{0cm}
\hspace*{-18cm}
\plotfiddle{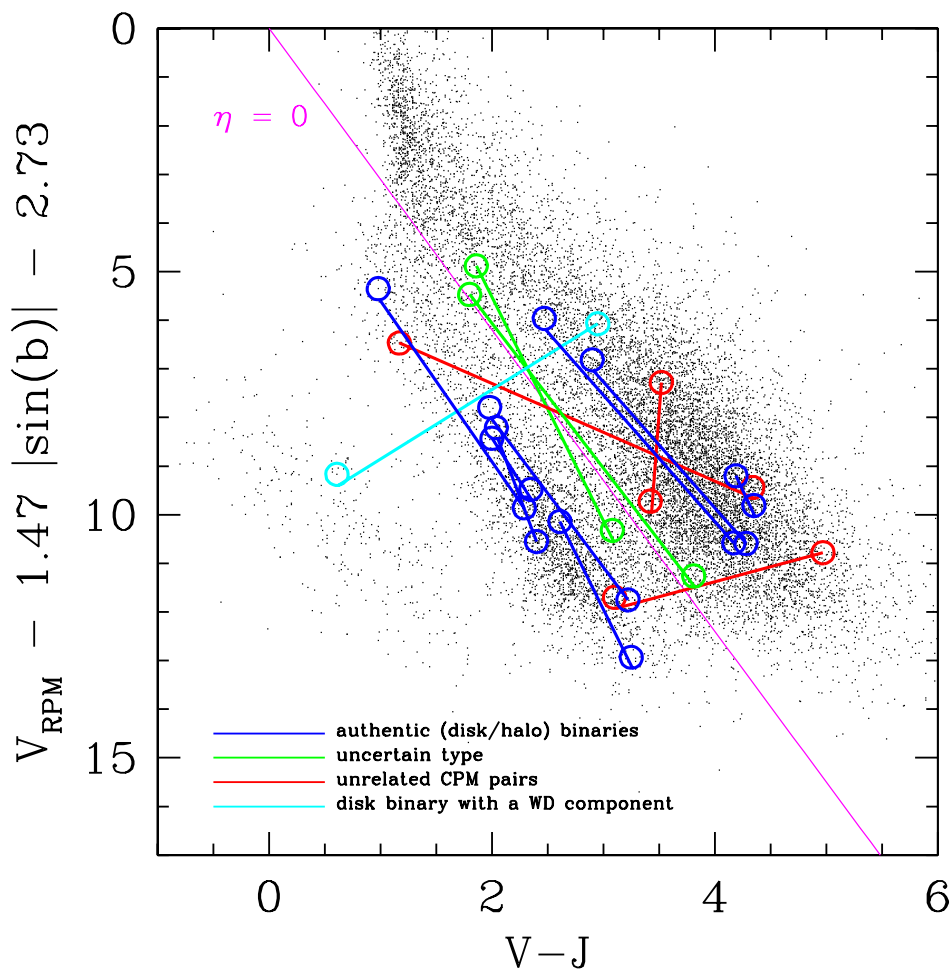}{1cm}{0}{0.7}{0.7}{0}{0}
\caption{\label{fig:two} RPM diagram adjusted for Galactic latitude
$b$ clearly separating MS, SD, and WD tracks, with $\eta=0$ (magenta
line) being our formal MS/SD discriminator.  A few representative
examples of how this diagram is used to classify binaries are shown by
pairs of circles connected by a line: genuine pairs, whether disk or
halo (blue); probably genuine but unclassifiable pairs (green);
optical pairs not physically associated (red); MS-WD pair (cyan).
Note that pairs that do not contain WDs are also regarded as
unphysical when the line connecting them is far from parallel to
either the MS or SD track.  }\end{figure}

When classifying the binaries, one expects not only to find that both
components belong to the same population, but also that their
positions on the RPM diagram lie on the same ``isotach''\footnote{An
isotach is essentially an isochrone vertically displaced in an amount
that is determined by the tangential velocity.}.  That is, since the
two members of a binary must have similar metallicities and
essentially the same proper motion but usually different luminosities,
one expects that the line connecting their positions on the RPM
diagram should be approximately parallel to the corresponding MS or SD
track for disk and halo binaries, respectively.  See figure 12 of
\cite{faint}.  Hence, by placing both components of the binary on an
RPM diagram, we not only classify it as disk or halo, but also subject
it to an extra test of the physical association of the two components.
The only cases permitted not to follow this ``parallel rule'' are
those involving a WD companion.  Binaries composed of one MS and one
SD member will be rejected as unrelated pairs.  So will binaries
composed of two MS or two SD stars if the line connecting them is
inconsistent (within measurement errors; typically, $\sigma_{\rm V}
\sim 0.25$ mag, see \citealt{bright} and \citealt{faint}) of being
parallel with the respective sequence.  In a very few cases, we find
that the connecting line straddles the disk/halo boundary, being
consistent within measurement errors of being either a disk or a halo
binary.  These binaries are excluded from the sample because this
ambiguity does not allow us to reliably classify them with the present
data.  Figure \ref{fig:two} illustrates various classification
examples, including good disk and SD binaries, cases of uncertain
type, as well as pairs rejected because of inconsistent membership.

Besides pairs with WD components, other possible cases of physical
pairs that would not necessarily follow the isotach criterion
described above will be those involving an evolved component, i.e.,
either stars at the turnoff or subgiants.  In practice, this is a
potential problem only for halo binaries since, as can be seen in the
RPM diagram in Figure \ref{fig:two}, at bright magnitudes the SD track
crosses the $\eta = 0$ boundary between disk and halo populations,
merging with the MS track.  Hence, a real halo binary with an evolved
component will likely fail our ``parallel rule''.  However, we can
easily account for this once the final vetting of disk/halo binaries
have been performed by searching among those pairs rejected by the
``parallel rule'' and that are composed by one star with a clear halo
classification as companion of a brighter star.  The results of this
search for evolved components is given at the end of \S 3.2.

Our philosophy is to effectively use all the information available for
each star in order to assign it to the disk or halo population, or
identify it as containing a WD.  According to the amount of
information available, we then construct several subgroups and study
them separately.  Just as during sample selection, we consider close
and wide binaries as independent subsets.  Inside any of these
subgroups, we typically must close-inspect several pairs to figure out
the source of either discrepant classifications or pairs containing MS
and SD members.  Whenever the source of the anomaly cannot be
determined, the pair is rejected.

The following subsections contain a detailed account of the
classification process for each subgroup of binaries and the final
tally of disk and halo memberships.  Some readers may wish to skip to
\S~\ref{sec:results}, where we present our results.

\subsection{Close Binaries
\label{sec:closeclass}}

Very close binaries are affected by blending, thereby corrupting the
RPMs, colors, and RPM discriminators $\eta$ of their components as
they are derived from rNLTT.  For example, suppose USNO-A could not
resolve a binary whose components have similar luminosities.  These
then would appear in rNLTT as having equal $V$ magnitudes, but too
bright by 0.75 mag.  Depending on whether or not the stars were also
unresolved in 2MASS, their colors could be incorrect as well, leading
to values of $\eta$ that could send one or both components to the
incorrect side of the RPM diagram and hence be either rejected as a
false match or erroneously classified.  With this in mind, we perform
a first subdivision of the close binaries into four subgroups
according to whether their components are resolved or blended by the
2MASS and USNO photometry.  Additionally, there are four other
subgroups according to whether the binary is resolved or not by 2MASS
and whether one or both of the binary's components do not have a match
in the USNO scanned plates.  Next we go one by one through these
subgroups and detail the classification scheme.

\noindent {\em 2MASS and USNO resolved binaries} [21]. These binaries
have all the possible information available, so no assumption has to
be made in their classification.  We compute $\eta$ for all the stars
using both the USNO-A photometry as well as the Luyten photometry and
obtain the same classification in every single case.  This is relevant
because it gives confidence for using Luyten photometry in the cases
for which USNO-A data are blended or not available.  One binary is
rejected because its components were classified as disk MS and halo
SD.

\noindent {\em 2MASS resolved and USNO blended binaries} [43]. These
binaries, although blended in USNO-A, were resolved by Luyten.
Prompted by the good results given by the Luyten photometry when the
USNO-A data are resolved (previous subgroup), we compute $\eta$ using
the resolved data from Luyten.  Additionally, we merge the 2MASS
resolved detections and, combine them with the blended USNO-A
measurement, and so compute $\eta$ for the ``merged binary''.  There
are only two cases for which the Luyten photometry and the merged
binary give different answers, and we choose to follow the former one.
There are no disk-halo cases in this subgroup.

\noindent {\em 2MASS blended and USNO resolved binaries} [1]. As 2MASS
has better angular resolution than USNO-A, these should be rare.
Indeed, the one example is a very special case.  The two components
are very faint, with Luyten magnitudes of $V\sim$ 19 and 18, and with
the brighter component being very blue both in the Luyten and USNO-A
colors, hence indicating a WD.

\noindent {\em 2MASS and USNO blended binaries} [127]. For these
binaries we compute $\eta$ for the merged binary, and compare it with
the value obtained from the 2MASS blended measurement and the optical
magnitude obtained by merging the resolved Luyten measurements.  In 11
of the cases ($\sim$ 9\%) we find disagreement between these
classifications.  We examine each of these individually in order to
determine the origin for the disagreement and so assign the most
appropriate classification.  There are also 9 cases of disk-halo
pairs, which are due to relatively large proper-motion differences and
the proximity of these binaries to the $\eta = 0$ boundary between the
two populations.

\noindent {\em 2MASS resolved and one component missing in USNO}
[98]. For this subgroup we compute $\eta$ for the component with
USNO-A data, and for both components using the resolved Luyten
photometry.  There are 11 cases for which both classifications for the
component having USNO-A data disagree and 13 cases of disk-halo pairs
from the Luyten classification.  Close inspection of the 11
disagreements favors the results from the Luyten photometry in all
cases, so these are adopted.  The 13 disk-halo pairs are rejected.

\noindent {\em 2MASS resolved and both components missing in USNO}
[53]. In these cases the only optical photometry available is
Luyten's, so we use it to compute $\eta$ for all components.  We find
only 4 cases are classified as disk-halo, and we reject them.

\noindent {\em 2MASS blended and one or both components missing in
USNO} [17]. Here the only alternative is to combine the optical light
given by the Luyten photometry and classify the merged binary as if it
were a single star.  Except for one binary with a probable WD
component, all the rest are classified as disk binaries.

\noindent {\em One component not in rNLTT} [34]. As with the cases
discussed in \S 2.2, we search for the companions of these rNLTT stars
in USNO-B, finding only 2 matches.  The remaining 32 binaries are
classified following the component present in rNLTT, and also
verifying that the NLTT colors are consistent with the components
lying on the same isotach.  In this way 15 are classified as disk
binaries, 4 as halo binaries, and 11 are disk pairs with a WD
component.  The remaining 2 binaries in this category have uncertain
classifications and are not included in the final samples.

\noindent {\em Neither component in rNLTT} [7]. These were found by
searching in the intersection between the Luyten notes and NLTT stars
that nominally lie in 2MASS areas but were not recovered by rNLTT.  We
accept only those pairs that have 2MASS measurements, whose 2MASS
separations and position angles are consistent with the Luyten notes,
and whose 2MASS entries do not have USNO-A counterparts, which would
indicate a slow-moving star.  Four of these 7 are considered real
binaries based on their RPMs, of which 2 contain a WD component.

While our philosophy is to accept all 393 close binaries
($\Delta\theta<10''$) as genuine, 24 systems could not be cleanly
classified and are removed from the sample.  Of the remaining 369
systems, 69 are classified as halo binaries, 278 as disk binaries, and
22 contain a WD companion.

\subsection{Wide Binaries
\label{sec:wideclass}}

Wide binaries are not as problematic as close binaries since they are
not usually affected by blending.  However, we still have to deal with
cases for which one or both components lack USNO-A and/or 2MASS data,
or for which one or both components are missing from rNLTT.  Moreover,
in contrast to the close binaries, we do not automatically accept the
wide binaries as genuine, but rather demand that both components sit
on the same isotach as described in \S~\ref{sec:classify}.

Whenever USNO-A photometry is not available, we derive the $V$
magnitude from Luyten photometry, which was shown to be a safe
procedure in the previous section (the 23 close binaries resolved in
both 2MASS and USNO yielded the same classifications regardless of the
use of USNO-A or Luyten photometry).  Of our 715 wide binaries with
both components present in rNLTT, 679 (90\% of entire wide sample)
have 2MASS data for both components, and hence their classification is
straightforward.  For the 30 wide binaries having one component
missing in 2MASS, we follow the disk/halo classification obtained from
the component with a near-IR measurement and check that the Luyten
photometry is consistent with the two stars lying approximately on the
same isotach.  Finally, 6 binaries lack 2MASS measurements for both
components but were confirmed as CPM systems in \S~\ref{sec:widegen}.
The 12 stars in these pairs are very bright Tycho-2 objects that
saturated the 2MASS detectors, and are all assigned to the disk
population.

In \S 2.2 we held for further inspection 4 pairs with large
proper-motion differences as estimated from the vector position
offsets of the two epochs because they were not found in USNO-B.  We
find here that 3 of them (NLTT 8502, 9308, 19207) have inconsistent
RPM positions (MS-SD pairs).  The fourth one (NLTT 34379), although
being an acceptable disk binary in the RPM diagram, is not recognized
as such by Luyten (i.e., non-Luyten pair).  Hence, since there is no
direct information on the actual proper-motion difference between the
components of this pair, we consider it an uncertain binary and do not
include it in the final clean sample.

As explained in previous sections, binaries with one component missing
from rNLTT are individually searched for in USNO-B.  Of the 43 wide
pairs in this category, the missing component of 23 are found in
USNO-B and easily classified.  Two of the binaries with rNLTT-missing
companions not found in USNO-B are then found in 2MASS and classified,
and another 6 that are part of very wide systems ($\Delta\theta >
200\arcsec$) are excluded because no information on their
proper-motion difference is available to judge their reality.  For
each of the remaining 12 binaries, we closely inspect the Luyten
photometry on both components, finding 5 pairs with a WD component and
confirming the other 7 as being approximately on the same isotach.  We
furthermore examine them in online (ESO database) images at separate
epochs, and visually confirm 2 pairs to be real CPM systems.  These 2
are classified following the type of the component present in rNLTT.
The remaining 5 pairs whose reality as CPM systems is difficult to
confirm from the images are excluded from the sample.

Next, given our goal to assemble a complete catalog of all the CPM
binaries present in NLTT, we also search for the subset of binaries
not recognized as such by Luyten (i.e., non-Luyten binaries) and that
have not been already found because one or both of its components were
not recovered by rNLTT.  As pointed out in \S 2.3, 138 of the 680 wide
candidate Luyten binaries had at least one component lacking an
independent proper-motion measurement.  In terms of wide binaries that
were accepted in the final samples, these last two numbers become 84
and 530, respectively.  Based on this experience with Luyten wide
binaries and scaling to the 44 wide non-Luyten binaries that we have
already accepted into our final clean samples, we estimate that there
are an additional $(44\times 84)/(530-84) \sim 8$ non-Luyten binaries
that we have missed.  In order to see if we can account for these, we
use USNO-B and perform a search around NLTT stars for CPM companions
that are also NLTT stars, expecting to find $\sim 6$ given the
incompleteness of USNO-B (which for binaries rises from 10\% at
30$\arcsec$ to 50\% at 10$\arcsec$; \citealt{gould03b}).  We find 3.
Of these, 2 turn out to be companions of binaries already in our
sample, making them triple systems, while the third one is a real pair
(NLTT 1024/1041) which was missed by us because rNLTT misidentified
the secondary, recovering instead an unrelated nearby object.  The
fact that we find 3 when we expected $\sim 6$ of these binaries is
mildly improbable ($\sim 15\%$).  However, the remaining $\sim 3$
missing binaries should have a negligible impact in our sample of
about a thousand binaries.

Finally, knowledge of radial velocities could potentially be of
importance for pairs classified as ``uncertain'' because their
components straddle the disk/halo boundary in the RPM diagram.
However, a search in SIMBAD for a randomly chosen third of these pairs
did not yield a single radial velocity.  Nevertheless, the final 41
CPM pairs in this category show a distribution of angular separations
consistent with that of the disk binaries for $\Delta\theta >
10\arcsec$ (see \S 4.3 below), and hence their exclusion from the
final samples does not affect our results and conclusions.

Of the $756-63=693$ wide CPM systems that passed through the selection
process of \S~\ref{sec:widegen}, 123 are excluded from the clean
sample by the classification procedure described above.  These 123
include 60 systems containing WD components, 58 that show either
unphysical MS-SD classification, inconsistent RPM positions, or are
considered of uncertain classification, and 5 for which one component
was not found in any available source catalog.  The final clean sample
consists of 570 wide binaries with solid classification: 523 belonging
to the disk population, and 47 belonging to the Galactic halo.  Figure
3 shows $\sim 10\%$ of our wide binaries, selected randomly from the
final clean samples.

At this point, we perform a search for real pairs that could have been
rejected by the classification procedure described above because of
the presence of an evolved component, as described in \S 3.  Recall
that this is a potential problem only for halo binaries, since it is
the SD track that crosses the disk/halo boundary of the RPM diagram at
bright magnitudes.  Among the 61 CPM pairs rejected because they
failed the ``parallel rule'' in the RPM diagram, we then search for
those involving a component classified as a halo star and a brighter
star whose position in the RPM diagram is consistent with it being a
turnoff star or a subgiant.  Among the 8 pairs so selected, 5 have
Hipparcos counterparts.  Of these, 2 have $M_{\rm V} \sim 4$ and $V-J
\sim 1$, i.e., consistent with being turnoff stars (NLTT 34611,
44964).  For the 3 pairs with no Hipparcos counterpart, we use the
component with halo classification to infer a photometric distance
(see \S 4.2) and with this distance estimate the companion's absolute
magnitude.  This produces one pair with a component having $M_{\rm V}
\sim 2.5$ and $V-J = 1.8$, consistent with being a subgiant star (NLTT
33701).  These 3 possible halo binaries with evolved components are
not considered part of our final halo sample because they cannot be
vetted as rigorously as other pairs in our sample.  Because the
possibility of a subgiant or turnoff component is not strongly
correlated with binary separation, the exclusion of these pairs does
not have any significant impact on the results we present in \S 4.3.

\section{Results
\label{sec:results}}

In this section we report on the properties of our final datasets.
Our main result consists of the determination of the binary frequency
as a function of the separation between the components.  This
distribution can be computed directly as a function of angular
separation on the sky by the simple counting of the number of binaries
in each interval of $\,\log\Delta\theta\,$.  However, since it is also
of interest to know the distribution of projected physical
separations, we need to assign distances to each of our binaries.
Furthermore, distances are needed in order to determine stellar masses
and study the distribution of mass-ratios between the components of
the binaries.  We accomplish this with the use of color-magnitude
relations (CMRs), derived separately for the disk and halo samples.
These, although not very precise as an absolute distance indicator for
any given system, can be used in a statistical way as a
characterization of the distances probed and to infer several overall
properties of this dataset.

In \S 4.1 we study the color distributions of the final samples and
compare them with the underlying rNLTT catalog.  In addition,
color-color diagrams are used to map the range of spectral types of
the stars in our datasets.  We then describe in \S 4.2 the derivation
of the CMRs that we use to determine individual distances to the
binaries, showing the distributions of distances and luminosity
functions of both the disk and halo samples.  The distributions of
angular and physical separations are presented in \S 4.3, and in \S
4.4 we show the distributions of luminosity and mass ratios between
the components of the binaries.

\subsection{Color distributions and spectral types
\label{sec:4.1}}

In Figure 4 we show the $V-J$ color distribution of the stars in
binaries in comparison with that of the entire rNLTT catalog, for disk
and halo populations separately.  The rNLTT disk and halo single stars
included in these plots are those whose membership to either one of
these types is very secure.  In terms of the discriminator $\eta$
introduced in \S 3, secure disk stars are those with $\eta\,< -1$, and
secure halo stars those with $1 < \eta < 4.15$, which produces sets of
17,690 and 4,883 stars, respectively.  The resulting color
distributions of both sets of binaries are, with a high degree of
confidence according to Kolmogorov-Smirnov goodness-of-fit tests,
different than those of the corresponding rNLTT stars.  These diagrams
show that both samples of binaries have larger fractions of bright
(i.e., blue) stars than the catalog as a whole.  This is a selection
effect due to the magnitude-limited nature of rNLTT (i.e., the
companions of bright stars are preferentially selected compared to
those of fainter ones), and should be kept in mind when interpreting
results regarding the distributions of luminosity and mass ratios.

In order to learn about their spectral characteristics, we plot in
Figure 5 color-color diagrams of the stars in our binaries.  Given the
proximity of these samples (see \S 4.2 below), reddening should have a
negligible effect on the main features of these diagrams.  The
approximate locations of dwarf stars of various spectral types are
indicated, as obtained from similar diagrams by \citet{giz00} and
\citet{fin00}.  Both plots indicate that the large majority of the
stars in these binaries are dwarfs of spectral types between M0 and M5
(M6 dwarfs have $J-K \gtrsim 1$).  The paucity of later-type stars is
almost certainly due to the confluence of the natural luminosity
function of M dwarfs and the fact that for disk stars, with distances
$d \sim 60$ pc, the catalog's $V \sim 19$ magnitude limit reduces the
effective volume sampled for $M_{\rm V} \sim 15$ \citep{faint}.  In
addition, Luyten may have had problems measuring the proper motions
for extremely red stars in the POSS plates: stars later than M5
usually fell below the ``O'' plate limits (T. Oswalt 2003, private
communication).  Less populated tails of late G and K stars complete
the datasets.  Almost all the stars have $V-J \gtrsim 1$,
corresponding to $M \lesssim 1\, M_{\odot}$ for disk MS stars and $M
\lesssim 0.8\, M_{\odot}$ for halo SDs.

\subsection{Disk and halo color-magnitude relations
\label{sec:4.2}}

Of the 801 disk binaries in our final sample, 242 have at least one
component catalogued in the Hipparcos database, 196 of which are wider
than $10\arcsec$.  In the case of the halo sample, 9 binaries have
Hipparcos parallaxes available, with 7 of them being wider than
$10\arcsec$.  However, most of these are bright primaries, and,
although their secondaries can be considered also as stars with
``measured'' parallaxes, they are not fully representative of either
the disk or halo samples of binaries.  For this reason we make use of
single stars in the Hipparcos database itself, as well as halo stars
present in rNLTT with parallaxes from other sources.  We thereby
assemble parallax samples covering the full color ranges spanned by
both our disk and halo binaries.

Figure 6 shows $(M_{V},V-J)$ color-magnitude diagrams (CMDs) for the
parallax samples and the polynomial fits we obtain from them.  The
small dots in the upper panel (6a) are all Hipparcos stars within 50
pc of the Sun, which very well cover all the blue half of our disk
sample.  These stars show a scatter with respect to the fit of 0.41
mag, consistent with that measured by \citet{reid91} for the solar
neighborhood.  The red half is filled with all the stars in the sample
of disk binaries that are companions to Hipparcos stars, and are shown
as open triangles.  The scatter with respect to the fit for the red
stars is 0.87 mag, much larger than that of the first group of stars.
While it is known that the scatter in the CMR depends on color and has
a maximum of 0.5 mag at $V-I \sim 2.5$ ($V-J \sim 4$)
\citep{reid91,mon92}, there are still 0.7 mag that cannot be explained
by the USNO-A photometric errors alone ($\sim 0.25$ mag).  We fit this
MS with a 7th-order polynomial, shown as the solid line, and use this
fit as our CMR for the disk sample.

Similarly, in Figure 6b we show the stars used to determine the halo
CMR.  The black dots are single halo stars present in rNLTT that have
Hipparcos parallaxes, and the open triangles are the components of the
9 halo binaries described before.  These two groups cover only the
blue half of our sample of halo binaries.  The red end of the SD track
is filled with 26 LHS stars (open circles) with parallaxes measured by
\citet{mon92} and \citet{giz97}, and that are classified as halo stars
using the rNLTT data.  We fit this halo CMR with a 5th-order
polynomial, shown as the solid line going through the data.  The
dashed line is the linear CMR obtained by Gould (2003) from a
kinematic analysis of 4588 SDs selected from rNLTT.

Figure 7 shows the CMD for all the disk and halo binaries with
separations larger than $10\arcsec$ (the wide sample) that
satisfactorily passed the selection and classification procedures
described in \S 2 and \S 3.  The solid lines are the CMRs derived from
the subsamples of rNLTT stars with parallaxes described above, and
placed at convenient distances that are discussed next.

Figure 8a shows the distribution of the distances to all our primaries
(in close and wide binaries) obtained with our fits to the CMRs.  Note
that the disk binaries are located at an average distance of 60 pc,
while the halo binaries lie farther away, with an average distance of
240 pc.  This is expected: since halo stars move faster than disk
stars, they can be detected at larger distances in a proper-motion
limited sample.  Finally, using these distances we plot in Figure 8b
the $V$-band luminosity functions (LF) for stars in disk and halo
binaries (both primaries and secondaries together).  The peak of the
disk LF occurs at $M_{\rm V} \sim 11$, which can be compared to the LF
of M dwarfs obtained from HST counts, which peaks between 11 and 12
\citep{zheng}.  In the case of the halo, the LF of binary components
appears to peak at $M_{\rm V}\sim 10$, very close to what Gould (2003)
found in his analysis of 4588 SDs selected from the same rNLTT.
Hence, whether in the disk or halo, the components of wide binaries in
our samples appear similar to single stars in the field population.

\subsection{Distributions of angular and projected physical separations
\label{sec:sep}}

The distributions of angular separations for our final samples of
(801) disk and (116) halo binaries are shown in Figure 9.  Recall that
our search extends to separations up to $500\arcsec$ for disk
binaries, and $900\arcsec$ for halo binaries.  The normalization is
set relative to the entire underlying rNLTT catalog, done separately
for disk and halo stars\footnote{The region of the rNLTT catalog given
  by the intersection of POSS I and 2MASS Second Incremental Release
  areas contains 20279 and 6834 disk and halo stars, respectively.}.
The error bars represent the Poisson errors ($N^{-1/2}/\ln 10$) in the
number of binaries falling into each bin of separation.

At close separations, $\Delta\theta < 10\arcsec$, selection effects
due to blending are dominant, and we are able to identify fewer and
fewer binaries as the separation decreases.  As discussed above, for
separations larger than 10$\arcsec$, both 2MASS and USNO, the sources
of rNLTT, are able to resolve even very bright stars and, furthermore,
our search strategy (\S 2) was implemented in such a way that we are
confident of being free of selection effects as a function of
separation (but see last paragraph in this same subsection).  At the
wide end, both distributions are well described by linear relations in
this log-log plot, corresponding to power laws of the form
$f(\Delta\theta) \propto (\Delta\theta)^{-\alpha}$.  Furthermore, the
slopes $\alpha$ of both wide distributions appear to be very similar.

In order to quantify this result, we fit both wide ends to the
functional form $f(\Delta\theta) = A\,(\Delta\theta)^{-\alpha}$.
Instead of fitting the binned data shown in Figure 9, a maximum
likelihood approach permits to use each binary as an independent data
point.  The disk distribution is fitted for the range $1.4 \lesssim
{\rm log}\,\Delta\theta \lesssim 2.7$ (from 25$\arcsec$ to
500$\arcsec$), which includes 323 binaries, and yields $\alpha = 1.67
\pm 0.07$.  The halo distribution is fitted for the range $0.74
\lesssim {\rm log}\,\Delta\theta \lesssim 2.95$ (from $5.\hskip-2pt
\arcsec 5$ to 900$\arcsec$), which includes 68 binaries, and yields
$\alpha = 1.55 \pm 0.10$.  Hence, the two slopes are consistent.  The
uncertainties in the slopes are determined analytically using equation
(2.4) of \citet{errslope}.

Next, in order to obtain the binary frequency as a function of
projected physical separation, distances to the binaries need to be
adopted.  For this, we first follow Figure 8a, and place all the disk
and halo binaries at their median distances of 60 and 240 pc,
respectively.  The resulting distributions of projected separation are
shown in Figure 10a, where the agreement between them can be clearly
appreciated.  Two things are particularly noteworthy.  First, with the
adopted distances, the two distributions match extremely well.  The
agreement between their normalizations is so good that the halo
distribution at its wide end appears just as a smooth continuation of
the trend of the disk binaries.  Second, both binary distributions cut
off exactly at the separations where we stopped searching for them,
{\em not showing any clear sign of a break or turnover} up to
projected physical separations of 0.1 pc and 1 pc, for the disk and
halo samples respectively.

If, instead of placing the disk and halo samples at their median
distances, we use the color-magnitude relations derived in \S 4.2 to
assign individual distances to each of the binaries, we obtain the
distributions shown in Figure 10b.\footnote{In Figure 10b we show the
actual counts in each bin of projected physical separation because,
since selection effects operate naturally on angular rather than
physical variables, its proper normalization is not a straightforward
procedure.  In particular, while on angular variables we can impose a
sharp cutoff in the search for CPM pairs and claim completeness as a
function of angular separation, when working on physical variables
this edge is smoothed and volume effects come into play.}  Note that
the distributions of binaries broaden, with some binaries going to
populate the region of small projected separations.  In addition, at
the wide end, the disk distribution now extends to physical
separations as large as the widest halo binaries.  Nevertheless, the
qualitative results are not changed and both distributions of binary
fraction are the same within the errors.

Finally, it must be mentioned that the flattening of the distribution
of disk binaries, occurring at $\Delta\theta \sim 10\arcsec -
25\arcsec$, is a puzzle to us.  It occurs well beyond the angular
separation at which blending is an issue, and it has no counterpart in
the angular distribution of halo binaries.  Indeed, this structure was
seen by \citet{garn91}, who states: ``The correlation shows a change
in slope around 25\arcsec. The index measured from the small angle
half of the data is $-0.96\pm 0.2$ while at large separations the
slope is significantly steeper, $-1.76.$'' However, since the
two-point correlation-function approach cannot be pushed to large
angular separations because of the rapidly increasing number of stars,
he interpreted the change in slope as a cutoff in the distribution.

We investigated the possibility that this flattening was due to a
hypothetical excess of bright stars present in the disk sample, but
introducing a cutoff in brightness did not remove it.  Finally, using
USNO-B we searched for all the non-NLTT CPM companions between
$10\arcsec$ and $30\arcsec$ of NLTT stars and found not nearly enough
companions to account for this deficit.  We discuss this further in \S
5.

\subsection{Luminosity and Mass Ratios
\label{sec:ratios}}

The left hand panels of Figure 11 (a,b) show the magnitude difference
(i.e., ratio of luminosities) between primaries and secondaries of
disk and halo binaries as a function of the apparent magnitude of the
primary in the V band.  We include only wide ($\Delta\theta >
10\arcsec$) binaries in the disk diagram, while for the halo diagram
we include binaries at all separations to increase the statistics.  At
any given brightness of the primary, there is a maximum observed ratio
of luminosities, occurring when secondaries are near the magnitude
limit of rNLTT ($V\sim 19$), which produces the upper envelope that
limits the location of the binaries in these diagrams.  Even in the
absence of a magnitude limit, the existence of the hydrogen burning
limit (M$\,\sim 0.08$ M$_{\odot}$) would set a very similar boundary
in plots like these.  In our case, at a mean distance of 60 pc, MS
stars of this limiting magnitude have masses slightly above 0.1
M$_{\odot}$, while SDs of this limiting magnitude at a mean distance
of 240 pc are of 0.2 M$_{\odot}$.  Due to this narrowing of the range
of magnitude differences as the primary gets fainter and fainter, it
would be misleading to construct a distribution of luminosity ratios
that includes the entire sample of binaries.  Instead, one should
study the luminosity ratios as a function of the luminosity of the
primary, i.e., the distribution of the luminosities of secondaries of
all binaries in a small interval of primary brightness.  To do this,
we select the three sections limited by four dashed vertical lines in
Figure 11a,b.  Inspection of the distribution of disk binaries in
these regions suggest a nearly uniform distribution of magnitude
differences all the way from zero (equal magnitudes) to the maximum
allowed value (secondary at the magnitude limit).  In the case of the
halo, it is possible to recognize a slight preference for smaller
magnitude differences, although the statistics are not as good as in
the disk case.

These raw distributions are, however, still incomplete descriptions of
the real situation, since they could possibly include many selection
biases.  For example, at any given primary brightness, it is
increasingly difficult to pick up the secondary as this gets fainter
and fainter, hence biasing the resulting distribution toward equal
luminosity components.  To account for this we replot the
distributions by normalizing the number of binaries in each bin with
respect to the relevant region of the rNLTT catalog (i.e., the
intersection between the POSS I and the 2MASS Second Incremental
Release).  To illustrate this normalization, consider a disk binary in
the brightest of the three selected regions, i.e., its primary having
a $V$ magnitude between 7 and 9.  Let's say $V_{1}=7.5$ and
$V_{2}=12$.  Then, the magnitude difference between this binary's
components falls in the bin $4 < \Delta\,V < 6$ of the distribution we
are building.  It is at this point where, instead of just adding
exactly one binary to this bin, we want to compare its secondary with
all the stars in the catalog that could have fallen in this same bin.
Hence, we count all the disk stars in rNLTT with $V$ magnitude between
$V_{1} + 4 = 11.5$ and $V_{1} + 6 = 13.5$, and use this number as the
normalization for the binary in question.  Finally, depending on
whether the samples of binaries are considered complete or incomplete
themselves, one can perform this normalization either considering or
not the completeness of rNLTT as a function of magnitude.  The
completeness function has been well characterized by \citet{gould03}
as part of his kinematic fit of halo parameters \footnote{The
completeness of rNLTT as a function of magnitude given in equation
(10) of \citet{gould03} has a typo in its third segment, $V_{\rm
break} < V < 20$.  The function should equal zero for $V = 20$.}.  The
center panels of Figure 11 (c,d) show the distributions of luminosity
ratios of disk and halo binaries normalized with respect to the raw
rNLTT catalog, while the right-hand panels (11e,f) show the same
distributions when correcting rNLTT for incompleteness.  The
single-star samples used for normalization are the secure disk and
halo ones introduced in \S 4.1.

To transform to mass ratios, we first obtain absolute magnitudes using
the CMRs found in \S 4.2, and then derive the masses from
mass-luminosity (ML) relations.  For the disk sample we use the
empirical ML relations of \citet{hmc93}, while for the halo sample we
use a theoretical ML relation corresponding to a metal-poor isochrone
of 10 Gyr.  This isochrone was built with the Yale Evolutionary Code
\citep{gue92}, running standard stellar models in steps of
0.05\,M$_{\odot}$ and metallicity ${\rm [Fe/H]} = -1.5$.  The
transformation from theoretical (M$_{\rm bol}$,T$_{\rm eff}$)
quantities to broad-band magnitudes and colors is performed using the
\citet{lej97} model atmospheres.

The results for mass ratios are presented in Figure 12, where we use
the same format as that of Figure 11 for the luminosity ratios.
Again, the disk diagrams include only wide ($\Delta\theta >
10\arcsec$) binaries, and the halo diagrams include binaries at all
separations to increase the statistics.  Figures 12a,b show the mass
ratio of each binary as a function of the mass of the primary.  The
vertical dashed lines indicate the regions chosen to compute the mass
ratio distributions that are shown in the central and right-hand
panels.  In Figure 12a, it is possible to see a slight
overconcentration of binaries that pile up to the right of M$_{\rm
primary} \sim 0.5\,{\rm M}_{\odot}$, which is due to a kink in the ML
relation of \citet{hmc93} at that same point.  Both {\it raw} (i.e.,
pre-normalized) distributions (Figs. 12a,b) show preference for equal
mass components.  This is most evident in the case of halo binaries,
but is also present in the disk binaries.

However, when normalized with respect to the entire rNLTT catalog
(Figs. 12c,d without correcting for incompleteness, and Figs. 12e,f
including such correction), a systematic pattern develops.  Except for
the solid-line histograms in the halo case (Figs. 12d,f), the first
two or three bins in all the distributions reveal a monotonic decrease
from equal-mass binaries toward systems of higher mass-ratio, which is
the expected behavior given the preference for equal-luminosity
components shown by Figure 11.  However, after these first bins, all
the normalized distributions are rising, eliminating any clear trend
in the overall distributions.  This is not expected given the
monotonic form of the distributions of luminosity-ratios, so is likely
to be due to some selection effect present in the data.

\section{Discussion 
\label{sec:discussion}}

\subsection{Power laws and limits on halo dark matter
\label{sec:machos}}

The most important result of this work is the measurement of the
distributions of angular and projected physical separations of disk
and halo wide binaries in the solar neighborhood.  We find that the
distributions of angular separations are well described by single
power laws over more than two decades of angular separation (Figs. 9
and 10).  Furthermore, the power laws extend all the way out to the
widest binaries in the samples, i.e., there is no evidence for a break
in either distribution up to projected physical separations of 0.1 pc
and 1 pc, for the disk and halo samples respectively.

These results have the potential to be very useful to impose
constraints on the nature of Galactic dark matter, as the widest
binaries, because of their small binding energies, are easier to
disrupt by passing encounters with massive objects.  For disk
binaries, issues like the existence of molecular clouds and spiral
arms, as well as the broad range of ages of the binaries themselves,
complicate any modeling of the interaction of the binaries and their
environment.  In the case of the Galactic halo, however, most of those
complexities are not present, and the results are easier to interpret.
A thorough investigation of the disruption of wide binaries is
presented in a companion paper \citep{jerry}, but a simple
order-of-magnitude calculation serves to illustrate this point.

Let us consider a halo binary of mass $m$ and semimajor axis $a$, and
suppose a black hole of mass $M$ with a velocity $v$ relative to the
binary passes by at a distance $b$ from the closest component of the
binary.  In the tidal limit ($b \gg a$), and using the impulse
approximation, the black hole induces a relative change of velocities
between the components of the binary given by,

\begin{equation}
|\Delta {\bf v_{1}}-\Delta {\bf v_{2}}| \equiv \Delta v_{12} = \frac{2\,G\,m}{v}\,\biggl(\frac{1}{b}-\frac{1}{b+a}\biggr) \simeq \frac{2\,G\,m}{b\,v}\,\frac{a}{b}
\end{equation}.

\noindent In order for this velocity change to disrupt the binary, we
require

\begin{equation}
(\Delta v_{12})^{2} \sim\,\frac{G\,m}{a}\,.
\end{equation}

\noindent In the tidal limit, binary disruption is dominated by the
perturber with the closest approach, which impact parameter can be
estimated from the rate equation $\pi b^{2} (\rho/M) v T = 1$, where
$\rho/M$ is the number density of black holes in the halo, and $T$ is
the binary's lifetime.  Replacing $b^{2}$ in the condition for
disruption, we obtain

\begin{equation}
a \sim \biggl(\frac{4\pi^2 G \rho^2 T^2}{m}\biggr)^{1/3} \sim\,0.1 \,{\rm pc}\,,
\end{equation}

\noindent where we have used $\rho \sim 0.009\,M_{\odot}\,{\rm
pc}^{-3}$, $T \sim$ 10 Gyr, and $m \sim 1\,M_{\odot}$.  This same
estimate is obtained by \citet{binney} for the disruptive effect of
molecular clouds on disk binaries.  Since the distribution of
separations for halo binaries (Fig 10) shows no signs of a break near
$a \sim\,0.1$ pc, one can infer that, if the Galactic halo is entirely
composed of black holes with a typical velocity of $v\sim 300\,\kms$,
then these cannot be more massive than $M \sim \pi a^2 \rho v T \sim
10^{3}\,M_{\odot}$.

The usefulness of this dataset to constrain halo dark matter relies
heavily on the genuineness of all the binaries comprising it, but most
importantly on those with the largest angular separations.  For this
reason we briefly discuss here the three widest halo binaries in our
sample, whose locations in the RPM diagram we show in Figure 13.  The
components of the pair NLTT 39456/39457 are both Hipparcos stars
showing consistent parallaxes, and have a proper-motion difference of
$3\,\masyr$ as measured by Tycho-2.  Their physical association is
beyond question.  The pair NLTT 16394/16407, a non-Luyten CPM pair and
the widest of all our binaries, has a proper-motion difference of
$9\,\masyr$ between its components and displays RPM positions nicely
following the track of the SDs.  Furthermore, it is confirmed by
USNO-B to be a CPM pair with $\Delta\mu = 6\,\masyr$, although both
stars have proper motions that are $20\,\masyr$ offset from the rNLTT
measurements.  However, this offset is not important since what
matters here is the CPM nature of the system being corroborated by two
independent catalogs, and hence we regard it as a real
physically-bound system.  The third CPM pair, NLTT 1715/1727, does
deserve a note of caution.  The proper-motion evidence, with
$\Delta\mu = 7\,\masyr$ as measured by rNLTT and which is confirmed by
USNO-B within $1\sigma$ of this measurement, argues for its reality.
However, although both of its components are clearly classified as
SDs, their relative position in the RPM diagram is not very
satisfactory.  As can be seen in Figure 13, NLTT 1715 and NLTT 1727
have essentially the same $(V-J)$ color, but their $V$-band
magnitudes, obtained from USNO-A, are different by $\Delta V \sim 1.5$
mag.  Moreover, these measurements are corroborated by the Luyten
photometry.  On the other hand, the error in the magnitude difference
is $\sigma_{\Delta V} = 2^{1/2}\times 0.25 \sim 0.4$ mag, is only
$1.25\sigma$ away from the 0.5 mag needed to put the RPM positions
parallel to the SD track.  Hence, it is possible that the
uncertainties in the photographic photometry are responsible for the
disagreement.  It is highly desirable then to obtain accurate optical
CCD photometry of these stars in order to settle this ambiguity.  A
search in SIMBAD for CCD photometry of these stars did not produce any
match.  We choose to include it in the final sample of halo binaries,
and in our companion paper \citep{jerry} we will address the impact of
its inclusion on the final results regarding limits on MACHO dark
matter.

\subsection{Disk versus halo binaries
\label{sec:diskvshalo}}

The second important aspect of this work is the comparison of the
properties of binaries that belong to different Galactic populations,
providing new insights on the early processes of star formation in
environments that are so radically different today.  We find, in the
first place, that the distributions of separations of wide binaries in
the disk and halo of the Galaxy are consistent with each other,
including both their slopes and normalizations.  Then, as shown in
Figure 11, there is also good agreement between disk and halo binaries
regarding their distributions of luminosity ratios: both populations
show an increasing number of binaries toward equal luminosity
components.  Finally, Figure 12 shows hints for a preference of
equal-mass binaries for both the disk and the halo (as one would
expect from the clear preference for equal luminosity components), but
the unexpected behavior of the distributions at large mass ratios
seems to indicate that low-mass stars are over represented in our
samples of binaries with respect to the underlying rNLTT catalog.
Nevertheless, although the possible presence of such selection effects
complicate their interpretation, both disk and halo distributions of
mass ratios are qualitatively similar.

All the above similarities between the two populations suggest that
stars in the disk and halo of the Galaxy formed under similar
conditions, kinematically cold enough to allow the formation of very
wide binaries in both populations.  Alternatively, if the star-forming
conditions in the early disk and halo of the Galaxy were very
different, one would be led to postulate the action of one or more
mechanisms, yet to be discovered, that make the properties of today's
disk and halo wide binaries as similar as we find in the present work.

There is, however, one puzzling difference between the disk and the
halo binaries, which was already noted in \S 4.3.  The distribution of
angular separations of disk binaries (Fig. 9) shows a break at
$\Delta\theta \sim 25\arcsec$, while no such feature is seen in the
distribution of halo binaries.  This break is not present in the
G-dwarf sample of \citet{duq91} or the M-dwarf sample of
\citet{fis92}, although their statistics are substantially smaller
than that of this work over the region of overlap.  However, it was
apparently seen by \citet{garn91} in his analysis based on correlation
functions (see end of \S 4.3).

Let us suppose, for a moment, that the origin of the peak is some
selection effect that has escaped our scrutiny.  Then we can think of
two possibilities.  First, perhaps blending is a much worse problem
than we estimated, and is removing binaries as wide as 20$\arcsec$
from our disk sample.  However, the distribution of halo binaries
strongly suggests that our sample is essentially complete down to
$\Delta\theta \sim 4\arcsec$, well below the $\Delta\theta =
10\arcsec$ separation that we conservatively adopt as a safe limit
above which one can regard our sample as free of selection effects due
to blending in the source catalogs.  Hence, this strange selection
effect would have to be removing more than a hundred disk binaries
while at the same time leaving the halo binaries almost untouched.  We
regard this hypothesis as very unlikely.  As a second possible
explanation, one might think that the presence of very bright stars in
the disk sample, which are mostly absent from the halo sample, is the
source of this irregularity.  To explore this possibility, we
progressively remove from the sample of disk binaries those having
bright components.  We find that no magnitude cutoff is able to make
the flattening disappear, and hence we also reject this second
explanation.

In an additional effort to understand this feature, we use USNO-B to
search the neighborhood of NLTT stars for non-NLTT CPM companions
between 10$\arcsec$ and 30$\arcsec$ (of course, also restricted to the
area of our survey, the intersection between POSS I and the 2MASS
Second Incremental Release).  This exercise has two advantages.
First, it serves the immediate purpose of studying whether the
flattening in the disk distribution of binaries is an artifact of the
NLTT catalog.  Second, since this search is restricted to a range of
separations narrow enough to allow us to ascertain the reality of the
candidates on a case-by-case basis (a rather painful task if we were
to search the entire range of separations of this catalog), we also
obtain an estimate of how many real binaries are being missed by
restricting ourselves to NLTT stars.

From inspection of Figure 9 one can see that $\sim 200$ binaries
between 10$\arcsec$ and 30$\arcsec$ are needed in order to make the
disk distribution consistent with the power law derived for
$\Delta\theta > 25\arcsec$.  The search in USNO-B returned 58
candidate CPM companions, of which 31 turn out to be due to the
diffraction spikes of bright NLTT stars, and 5 are USNO-B
misidentifications of the NLTT star.  Thus, we find 22 real non-NLTT
CPM companions of NLTT stars in the range of separations between
10$\arcsec$ and 30$\arcsec$.  The magnitude distribution of these
stars is presented in Figure 14, where we compare it with that of the
secondaries of disk binaries in our final sample.  Most of the newly
found CPM companions are indeed very faint, corresponding exactly with
the secondaries we know we must be missing due to the declining NLTT
completeness at these faint magnitudes \citep{gould03}.  In
conclusion, this search outside NLTT essentially added only those
binaries already expected to be missing because of NLTT's magnitude
limit, and these account for only $\sim 10\%$ of the extra binaries
needed to explain the flattening in the disk distribution as an
artifact of NLTT.

We cannot think of any other obvious selection effect that could be
responsible for this peak in the disk distribution of angular
separations while at the same time not producing a similar feature in
the corresponding distribution of halo binaries.  In terms of
projected physical separations, Figure 10 shows the flattening in the
disk distribution occurring at $r_{\perp} \sim 1,500$ AU.  Since the
power law of the halo distribution does not probe too far inside this
region of projected separation (there may be a hint that it goes to
$r_{\perp} \sim 1,000$ AU, but this is already the region where
blending begins to affect our completeness), we cannot determine
whether the same feature is present or not in the distribution of halo
projected physical separations.

Finally, a few words must be said regarding the potential effect that
very close companions (i.e., $a \lesssim 100$ AU) could have on our
results.  The identification of binaries based on CPMs is limited to
stars with proper motions large enough to be detected over the given
temporal baseline, but also to systems wide enough to be resolved in
the images of at least one of the two epochs.  A non-negligible
fraction of our stars, both primaries and secondaries, should
certainly have close companions that elude our detection (see the
high-resolution imaging survey of late-type stars of
\citealt{close03}, and the recent detection of a brown dwarf companion
to the Luyten star LHS 2397a by \citealt{melanie}), and these raise
the reasonable concern of whether our results regarding luminosity
and/or mass ratios could be significantly affected.  Based on
\citet{fis92}, about 35\% of M stars have a companion inside 100 AU,
which is closer than typically probed in our sample.  Hence, the light
of such close companions is included in the photometry of a similar
fraction of our primaries and secondaries.  We recall here that the
typical error in the $V$ photometry used in this work is $\sim 0.25$
mag, which translates to an uncertainty of $\sim 0.6$ mag in $M_V$
(before accounting for the intrinsic scatter in the CMR), and to more
than 0.7 mag in the RPM.  Hence, taking into account that companions
typically are significantly less massive than their primaries, their
contribution to the photometry of our stars is a minor issue, one that
certainly gets erased by the uncertainties of the available data.

\subsection{Comparison to the samples of \citet{ryan92} and \citet{allen}
\label{sec:allen}}

Selecting CPM pairs with RPMs indicative of halo stars, \citet{ryan92}
constructed a sample of 25 wide ($\Delta\theta > 10\arcsec$) halo
binaries from NLTT.  Of these 25, 2 are actually part of triple
systems and 10 are outside the intersection of POSS I and the 2MASS
Second Incremental Release, leaving 13 CPM pairs inside the area of
our survey.  We indeed recover these 13 binaries, although 4 of them
(with primaries NLTT 9308, 13320, 16468, and 54708) we classify as
disk binaries, while a fifth one (NLTT 28236) is not included in our
final samples because of a large proper-motion difference between its
components.  Recall that the selection and classification based on
rNLTT data have several advantages with respect to that based on NLTT
data alone, most notably, the availability of independent
proper-motion measurements for the components as well as a larger
color baseline (optical-infrared color).  Hence, there is agreement in
the classification of 8 of the 13 wide halo binaries common to our
sample.  Finally, \citet{ryan92} does not attempt to study the
distribution of angular separations with this small sample, and
focuses instead on a discussion of photometric parallaxes and the
fraction of close companions to some of the CPM stars.

\citet{allen} searched in NLTT for CPM companions of a sample of local
high-velocity and metal-poor stars with Str$\ddot{\rm o}$mgren
photometry.  They compiled a total of 122 wide binaries ($a >$ 25 AU)
and, by computing their Galactic orbits along with their photometric
metallicities, classified them as disk, thick disk, or halo stars.
Since their underlying source for identifying binaries is the NLTT
catalog, we compare their sample with that presented in this work and
contrast our results.  Note that, since Luyten intentionally recorded
as identical the proper motions of the components of systems he
regarded as binaries (this, despite the fact that he was in some cases
able to measure their actual non-zero difference in proper motion),
\citet{allen} have no way to independently measure the proper-motion
difference between the components and hence to identify and exclude
false associations from their sample, as we do with our binaries in \S
2.

Of their 122 binaries, 55 match the conditions of our search (that is,
stars inside the intersection of POSS I and the 2MASS Second
Incremental Release), of which 37 are common to our final clean
samples.  Examination of these 37 common binaries in the RPM diagram
of Figure 15 reveals that almost all of them have G-type primaries
(compare the binaries in Fig. 15 with those randomly selected from our
final samples and shown in Fig. 3).  Of the remaining $55-37=18$
binaries, 4 are actually in our datasets but were rejected because of
various reasons: one is actually a triple system, another includes a
WD companion, and the other two have components with inconsistent
positions on the RPM diagram (these last two pairs can be seen in
Fig. 15 as the lines crossing the $\eta=0$ boundary).  This leaves 14
binaries in the \citet{allen} sample that were not picked up by our
search.  However, all these 14 systems have angular separations
smaller than 5$\arcsec$, which is exactly the region where blending of
the images of the two components limits our completeness.  In
conclusion, the \citet{allen} sample supports our claim that we have
recovered essentially all the CPM binaries with separations larger
than 10$\arcsec$.

While we classify our binaries as belonging to either the Galactic
disk or halo on the basis of the RPM of the components (\S 3),
\citet{allen} use the complete space velocity of their stars (to
compute Galactic orbits) together with the metallicity estimate from
Str$\ddot{\rm o}$mgren photometry to assign their binaries to a given
population.  Of the 37 binaries common to both works, there is perfect
agreement in the classifications for all but 3 binaries.  Given the
fact that \citet{allen} have more information available to classify
their binaries, this agreement shows that our classification scheme
based on RPM diagrams works very well.  Regarding the 3 discrepant
cases (HIP 25137, HIP 85378, and G85$-17$), they classify them as halo
binaries, while we say disk.  The space velocities of these 3 binaries
would be consistent with either thick-disk or halo kinematics.  The
metallicities, [Fe/H]$\sim -0.6$, are however perfectly consistent
with the thick-disk, but would lie at the high end of the halo
metallicity distribution.  Hence, we believe it would be appropriate
to classify them as thick-disk stars.  In any case, and despite these
3 discrepant cases, the agreement in the classification of the set of
binaries common to both works is very good.

Although not directly mining the entire catalog for a complete sample
of CPM binaries, the work of \citet{allen} constitutes the most
extensive attempt at studying NLTT binaries previous to the present
work.  One of their main conclusions is that the separations of wide
binaries follow Oepik's (1924) distribution ($f(\Delta\theta)\propto
\Delta\theta^{-1}$), which does not agree with the power laws derived
in this work.  The reason for this disagreement is probably twofold.
On the one hand, they do not take into account the problem of blending
at close angular separations, which causes them to miss (just as we do
in our own samples) binaries with faint secondaries that are too close
to a brighter primary.  Second, they choose to visualize their data
using cumulative distributions, which, in complicity with the first
issue, are not as straightforward to interpret as the differential
log-log distributions that we use in Figures 9 and 10.

In Figure 16 we compare the \citet{allen} data with ours, both as
cumulative distributions (upper panel) as well as in differential
log-log form (lower panel).  In the cumulative distributions we only
compare binaries in the range 10$\arcsec$ to 500$\arcsec$ , where we
know the problems of blending are negligible and both samples should
be essentially free of selection effects as a function of separation.
Oepik's law is represented by the straight dashed line joining the
first and last data points, and it is immediately obvious that their
sample (represented as crosses) does not follow it.  Instead, their
cumulative distribution for $10\arcsec < \Delta\theta < 500\arcsec$
looks very similar to that of our sample of wide disk binaries (solid
line), which was shown in \S 4.3 to follow a power law with slope of
$\alpha = -1.67$, considerably steeper than Oepik's distribution.
Finally, the lower panel of Figure 16 shows that the differential
log-log distributions of the entire \citet{allen} sample (i.e., not
restricted to $\Delta\theta > 10\arcsec$) and our sample of disk
binaries are in almost perfect agreement for $\Delta\theta \gtrsim
25\arcsec$.  Hence, one can appreciate how much would the
\citet{allen} conclusions have changed if they had used differential
rather than cumulative distributions to visualize their data.

\section{Conclusions
\label{sec:conclusions}}

We have compiled a catalog of wide binaries ($a \gtrsim 100$ AU)
selected from among all the common proper motion (CPM) systems present
in the rNLTT catalog, restricting the search to the area of the sky
comprised by the intersection between POSS I and the 2MASS Second
Incremental Release.  With the help of the recently released USNO-B
catalog, our search has been extended to most NLTT stars in this
overlap region that were not initially recovered by rNLTT.  Given that
independent proper motions are available for essentially all the
components of our CPM pairs, we impose a selection criterion based on
the proper-motion difference as a function of angular separation
($\Delta\mu - \Delta\theta$ plane) to select good binary candidates.
The selected sample is then classified with the aid of reduced proper
motion (RPM) diagrams into disk and halo subsamples.  The
classification procedure also serves to identify unrelated pairs that
were not rejected by the selection cut (mostly, unphysical pairs with
disk and halo components at the same time).  Disk binaries are
searched up to angular separations of 500$\arcsec$, and halo binaries
up to 900$\arcsec$.  These correspond to projected physical
separations of 0.1 pc and 1 pc, for the disk and halo samples
respectively.

The final clean samples have 801 and 116 disk and halo binaries
respectively, by far the largest dataset of wide binaries available to
date.  The subsets restricted to separations larger than $10\arcsec$,
with 523 and 47 disk and halo binaries respectively, are essentially
free of selection effects as a function of angular separation.  Most
of the stars in these binaries have spectral types between M0 and M5.
The catalog also includes 82 CPM pairs with WD components.  All of
these pairs are classified as disk binaries, with one exception, NLTT
307/308.  The components of this pair have positions in the RPM
diagram consistent with a halo star and a WD, but their physical
association, based on their proper motions, is not clear cut, and it
should be treated with caution.

We compute the distributions of angular and physical separations of
the final samples of disk and halo binaries (Figs. 9 and 10).  Both
distributions follow power laws of the form $dN/d\Delta\theta \propto
(\Delta\theta)^{-\alpha}$, and we find $\alpha=1.67\pm0.07$ and
$\alpha=1.55\pm 0.10$ for the disk and halo respectively.  Hence, the
two slopes are consistent at the $1\sigma$ level.  Furthermore, their
normalizations are also consistent.  We also compute the distributions
of luminosity and mass ratios between primaries and secondaries
(Figs. 11 and 12), and find that disk and halo binaries are also
similar in these respects: both binary populations show a clear
preference for equal-luminosity components, and a somewhat less clear
(probably because of selection effects) preference for equal-mass
components.

In light of all these similarities we conclude that, despite the fact
that the disk and halo binaries belong to very different stellar
populations today, they probably formed under similar environmental
conditions, kinematically cold enough to produce bound systems as wide
as those reported here.  At least as concerns binaries, the end
result of the star-formation process on large scales seems to be
independent of the metallicity of the environment.

We find that the distribution of disk binaries flattens in the region
of $10\arcsec - 25\arcsec$.  Though unexpected, this feature has a
high statistical significance, and occurs in a range of angular
separation where selection effects due to blending are not at work.
We have explored various scenarios that could hypothetically account
for this flattening, even looking for more CPM companions outside
NLTT, but nothing could remove it.  Hence, we consider this flattening
a real structure in the distribution of wide disk binaries.  Since
they are farther away than the disk sample, the halo binaries do not
probe the same range of physical separation, and we therefore cannot
explore whether the same structure is present or not in the halo
distribution.

Finally, both disk and halo binaries show no evidence for a break or
turnover in their distributions, smoothly extending all the way up to
the limits of our search.  Given that the widest binaries are easily
disrupted by close encounters with large mass concentrations, these
results provide the opportunity to place limits on the nature and
properties of dark matter in the Galactic halo.  In a companion paper,
\citet{jerry} make a detailed investigation of these limits.

\begin{acknowledgements}

We thank D. Monet and the USNO-B team for providing us with a copy of
the USNO-B1.0 catalog.  We would also like to thank the referee, Terry
Oswalt, whose comments and suggestions led to substantial improvements
of the paper.  This publication has made use of catalogs from the
Astronomical Data Center at NASA Goddard Space Flight Center, VizieR
and SIMBAD databases operated at CDS, Strasbourg, France, and data
products from the Two Micron All Sky Survey, which is a joint project
of the University of Massachusetts and the IPAC/Caltech, funded by
NASA and the NSF.  The ESO Online Digitized Sky Survey is based on
photographic data obtained using The UK Schmidt Telescope. The UK
Schmidt Telescope was operated by the Royal Observatory Edinburgh,
with funding from the UK Science and Engineering Research Council,
until 1988 June, and thereafter by the Anglo-Australian
Observatory. Original plate material is copyright (c) of the Royal
Observatory Edinburgh and the Anglo-Australian Observatory. The plates
were processed into the present compressed digital form with their
permission. The Digitized Sky Survey was produced at the Space
Telescope Science Institute under US Government grant NAG W-2166.
This work was supported in part by grant AST 02-01266 from the NSF and
by JPL contract 1226901.

\end{acknowledgements}

\appendix

\section{Description of Catalog}

The catalog is arranged so that each line describes one pair of stars.
The data for each pair is grouped in six sections, 1) identifiers, 2)
positions, 3) proper motions, 4) photometry, 5) 3-digit source codes,
6) binary information.  Sections 1 to 5 give the corresponding
information for the two stars immediately next to each other.  That
is, section 3, for example, includes 4 entries: 2 for the two
components of the proper motion of the first star, and 2 for the
components of the second star.

The ordering of the two stars in each pair follows the corresponding
NLTT numbers in ascending order, and from this point on, the first
star in each pair will be labeled A, and the second will be labeled B.
Section 1 contains two entries: 1) NLTT(A), 2) NLTT(B).  Section 2
contains 4 entries: 3) $\alpha$(A), 4) $\delta$(A), 5) $\alpha$(B), 6)
$\delta$(B), where all coordinates are epoch and equinox 2000.
Section 3 contains: 7) $\mu_{\alpha}$(A), 8) $\mu_{\delta}$(A), 9)
$\mu_{\alpha}$(B), 10) $\mu_{\delta}$(B), in units of arcsec$\,{\rm
yr^{-1}}$.  Section 4 contains the photometry: 11) $V_{\rm A}$, 12)
$(V-J)_{\rm A}$, 13) $V_{\rm B}$, 14) $(V-J)_{\rm B}$.  The
uncertainties in the $V$-band photometry are dependent on the source
catalog: $\sim 0.25$ mag for USNO photometry, and $\sim 0.01$ mag for
Hipparcos.  The errors in the color are essentially those of the
optical data.  Section 5 contains two entries: 15) 3-digit source code
for A, 16) 3-digit source code for B.

The three digits of the source code are the same as those introduced
in \citet{faint}.  As summarized there, the digits refer to the
sources of the position, proper motion, and $V$ photometry.  1 =
Hipparcos, 2 = Tycho-2, 3 = Tycho Double Star Catalog, 4 = Starnet, 5
= USNO/2MASS, 6 = NLTT, 7 = USNO (for position) or common proper
motion companion (for proper motion).  More specifically, ``555''
means 2MASS based position, USNO based $V$ photometry, and USNO/2MASS
based proper motion.  For additional details see \citet{faint}.  For
the purposes of this catalog, we introduce two more digits for coding:
0 = the star is not recovered by rNLTT, and the position is the same
as that of the rNLTT companion, 8 = USNO-B.

Finally, section 6 of each catalog line contains the binary
information: 17) classification code for the common proper-motion
pair, 18) magnitude of the vector proper-motion difference between the
components (arcsec$\,{\rm yr^{-1}}$), 19) angular separation (arcsec),
20) position angle of B with respect to A (degrees), 21) rNLTT
binarity indicator, 22) indicates whether the pair is (1) or is not
(0) in the sample of \citet{allen}.  The classification code indicates:
1 = disk binary, 2 = halo binary, 3 = at least one component is a
white dwarf, 4 = rejected because of uncertain classification, 5 =
rejected because of components have inconsistent RPM positions, 6 =
rejected because of a large proper-motion difference or it is beyond
the limits of our search, 7) rejected because one component was not
found in any available sources.  Regarding the binarity indicator: 0 =
both NLTT stars are not present in rNLTT, 1 = NLTT regards it as a
binary but rNLTT did not recover one of the components, 2 = NLTT
regards it as a binary and both components are present in rNLTT, 3 =
NLTT does not regard this as a binary.

The catalog in ASCII format can be retrieved from\\
http://www.astronomy.ohio-state.edu/$\sim$gould/rNLTT$\_$binaries/binaries.dat.gz.
\notetoeditor{Previously ApJ had problems with interpreting urls
that contain tilde. For reference, the address should have a slash,
then tilde in front of gould. No spaces} 

The Fortran format statement
for the catalog record is:\\
(i5,i6,4f11.6,4f8.4,4f6.2,1x,3i1,1x,3i1,i2,f8.4,f7.1,f6.1,2i2)

\section{New halo stars with Hipparcos parallaxes}

In Table 2 we present 11 halo binaries for which one of the components
is an Hipparcos star, and hence, their companions become new halo
stars with available parallaxes.  Of the 11 binaries, only 5 have
accurate parallaxes, and these correspond to the first entries of the
table.

The fields in Table 2 are as follows: 1) Hipparcos ID, 2) NLTT number
of the Hipparcos star, 3) NLTT number of the companion of the
Hipparcos star (i.e., the new halo star with parallax).  From this
point, the Hipparcos star is labeled A, and the companion is labeled
B.  The next entries are: 4) angular separation (arcsec), 5) position
angle of B with respect to A, 6) $M_{\rm V}$(A), 7) $V_{\rm A}$, 8)
$(V-J)_{\rm A}$, 9) $M_{\rm V}$(B), 10) $V_{\rm B}$, 11) $(V-J)_{\rm
B}$, 12) parallax (mas), 13) parallax uncertainty (mas).  A
superscript 'a' indicates three binaries for which one of the stars
straddle the boundary between disk and halo stars, and should not be
taken as halo binaries with full confidence.

\begin{deluxetable}{rrrrrrrrrrrrr}
\footnotesize
\tablewidth{0pc}
\tablecaption{Companions to Hipparcos Halo Stars}
\tablehead{
\colhead{Hip.~\#}  
& \colhead{NLTT}  
& \colhead{NLTT}  
& \colhead{$\Delta\theta$}  
& \colhead{p.a.}
& \colhead{$M_V$}
& \colhead{$V$}
& \colhead{$V$-$J$}
& \colhead{$M_V$}
& \colhead{$V$}
& \colhead{$V$-$J$}
& \colhead{$\pi$}
& \colhead{$\sigma(\pi)$}
\\
\colhead{} & 
\colhead{(1)} & 
\colhead{(2)} & 
\colhead{$''$} & 
\colhead{deg} & 
\colhead{(1)} & 
\colhead{(1)} & 
\colhead{(1)} & 
\colhead{(2)} & 
\colhead{(2)} & 
\colhead{(2)} & 
& 
}
\startdata

  3187$^a$ &  2163 &  2167 &  27.7 &  19.5 &  4.8 &  9.88 & 1.19 & 14.0 & 19.04 & 4.95 &  9.70 &  1.30\\
 15126$^a$ & 10356 & 10349 &  78.4 & 132.7 &  5.7 & 10.23 & 1.33 & 10.3 & 14.83 & 3.21 & 12.64 &  1.66\\
 40068     & 18931 & 18924 & 110.4 & 208.5 &  3.0 & 10.01 & 1.31 &  7.7 & 14.75 & 2.22 &  3.91 &  1.22\\
 43490     & 20570 & 20571 &  37.3 & 182.3 &  5.3 &  9.55 & 1.16 &  7.2 & 11.46 & 1.72 & 14.10 &  0.91\\
 89523$^a$ & 46270 & 46279 & 110.6 &  32.5 &  5.5 & 10.13 & 1.36 & 10.7 & 15.31 & 3.78 & 12.04 &  1.07\\
\\
   911     &   525 &   526 &   8.7 & 168.9 & -9.0 & 11.80 & 0.97 & -9.0 & 14.35 & 2.16 &  6.13 &  5.67\\
 15396     & 10536 & 10548 & 185.7 &  85.5 & -9.0 & 11.22 & 0.98 & -9.0 & 15.78 & 2.29 &  3.78 &  2.27\\
 16683     & 11300 & 11288 & 223.5 & 263.6 & -9.0 & 11.42 & 1.18 & -9.0 & 14.64 & 2.15 &  5.89 &  2.77\\
 52854     & 25404 & 25403 &  25.9 & 296.0 & -9.0 & 11.43 & 1.19 & -9.0 & 13.01 & 1.69 & 10.25 &  7.79\\
 60849     & 30838 & 30837 &  15.7 &  23.9 & -9.0 & 12.55 & 1.13 & -9.0 & 19.09 & 3.03 & -3.11 &  5.66\\
 65418     & 34019 & 33984 & 490.6 &  84.4 & -9.0 & 12.18 & 1.07 & -9.0 & 16.10 & 2.26 &  2.52 &  3.57\\

\enddata

\tablecomments{a = indicates binaries that straddle disk-halo boundary   
}

\end{deluxetable}

\clearpage

\begin{figure}
\vspace*{0cm}
\hspace*{-18cm}
\plotfiddle{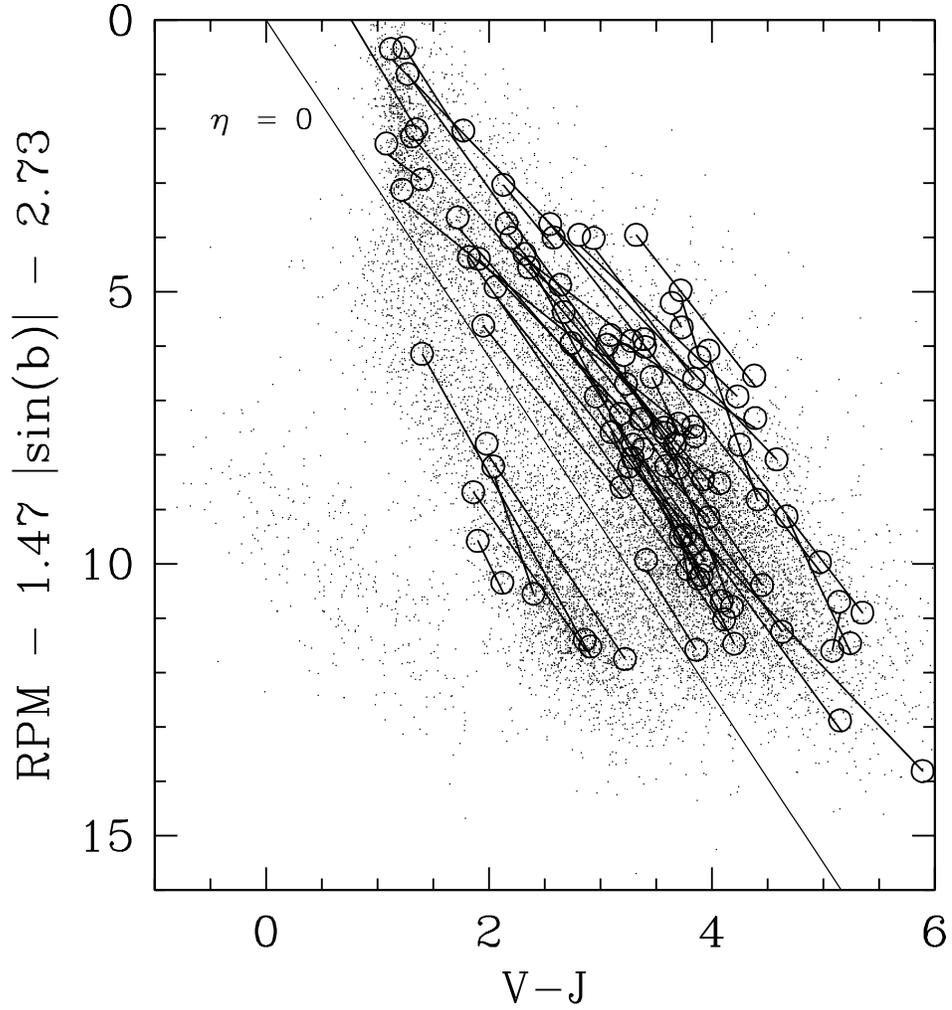}{1cm}{0}{0.7}{0.7}{0}{0}
\caption{\label{fig:three} Graphical representation of $\sim 10\%$ of
our binaries, randomly selected from the final clean wide sample
($\Delta\theta > 10\arcsec$).  Note how real binaries have their
components lying on the same isotach, i.e., the line connecting them
being approximately parallel to either the MS or SD track.
}\end{figure}

\begin{figure}
\vspace*{0cm}
\hspace*{-17cm}
\plotfiddle{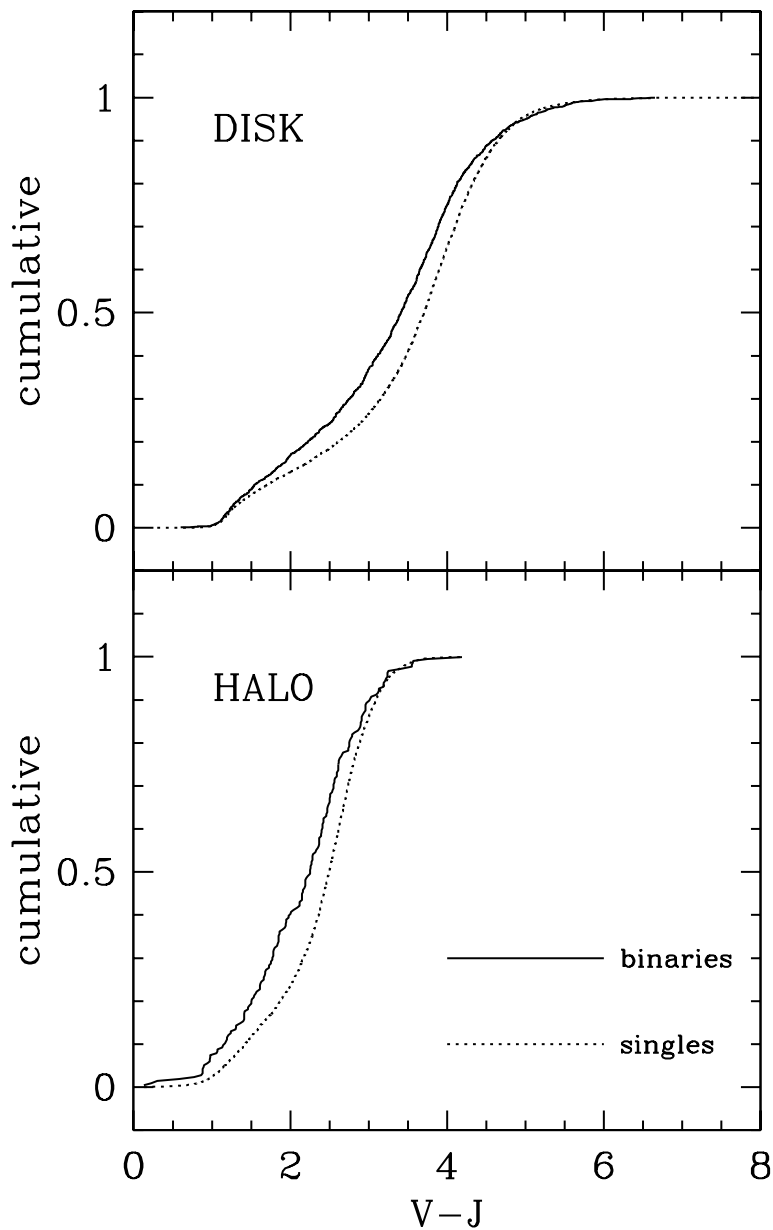}{1cm}{0}{0.75}{0.75}{0}{0}
\vspace{-2.0cm}
\caption{\label{fig:four}
Cumulative color distribution of the stars in binaries compared to that
of the entire rNLTT catalog, for the disk and halo populations separately.
Due to the magnitude-limited nature of the rNLTT catalog, the 
companions of bright stars are preferentially selected in comparison to
those of fainter ones, causing both samples of binaries to have larger
fractions of blue (i.e., bright) stars than the catalog as a whole.
}\end{figure}

\begin{figure}
\vspace*{0cm}
\hspace*{-18cm}
\plotfiddle{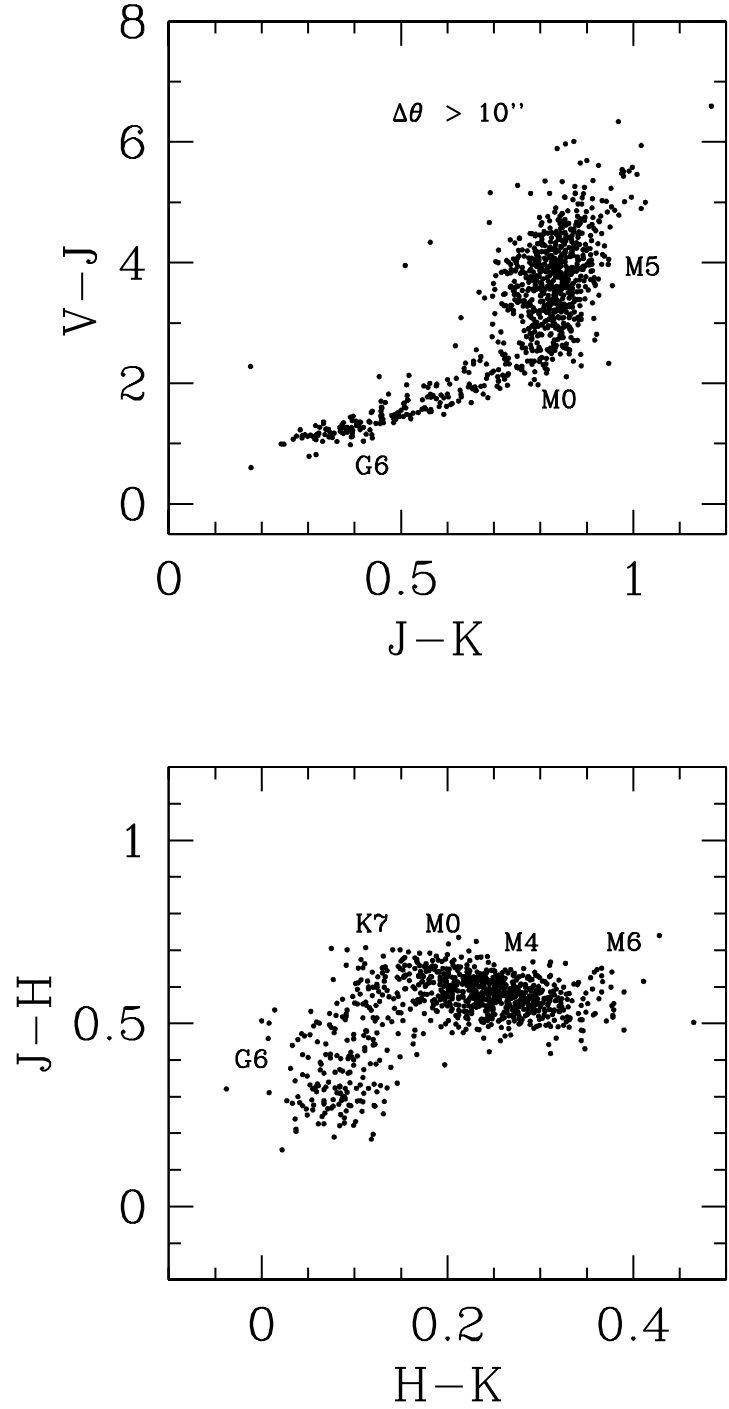}{1cm}{0}{0.75}{0.75}{0}{0}
\caption{\label{fig:five}
Color-color diagrams for all stars in the wide ($\Delta\theta > 10\arcsec$) 
sample, showing the approximate location of representative types of
dwarf stars.  Most of the stars in our binaries are of types M0 to M5.
}\end{figure}

\begin{figure}
\vspace*{0cm}
\hspace*{-17cm}
\plotfiddle{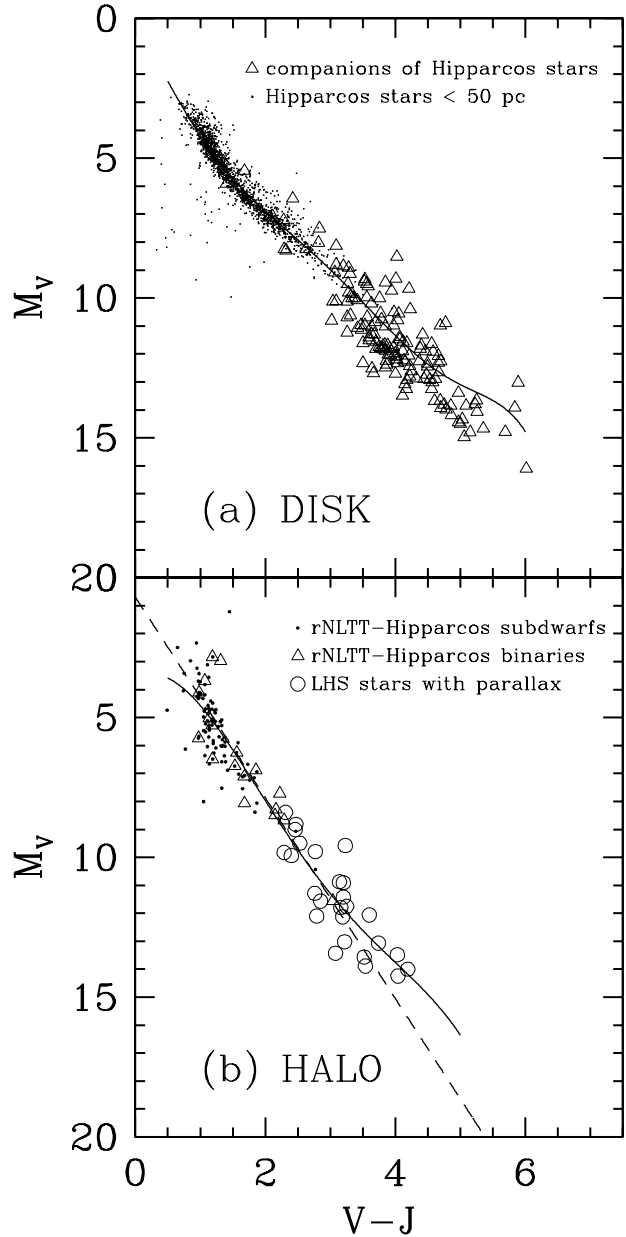}{1cm}{0}{0.75}{0.75}{0}{0}
\vspace{-2.0cm}
\caption{\label{fig:six} Color-magnitude diagrams of the parallax
samples used in the determination of color-magnitude relations.  (a)
Disk parallax sample: small dots are all the Hipparcos stars within 50
pc of the Sun in the color range of interest.  Open triangles
represent all the companions of stars in our disk sample that have an
Hipparcos parallax.  (b) Halo parallax sample: small dots are all SDs
present in rNLTT with Hipparcos parallaxes.  Open triangles are all
the rNLTT stars in binaries that have one component with an Hipparcos
parallax.  Open circles are SDs with parallaxes from \citet{giz97} and
\citet{mon92}.  The solid lines are the polynomial fits to these data.
The dashed line in the halo diagram is the linear fit from the
kinematic analysis of \citet{gould03}.  }\end{figure}

\begin{figure}
\plotone{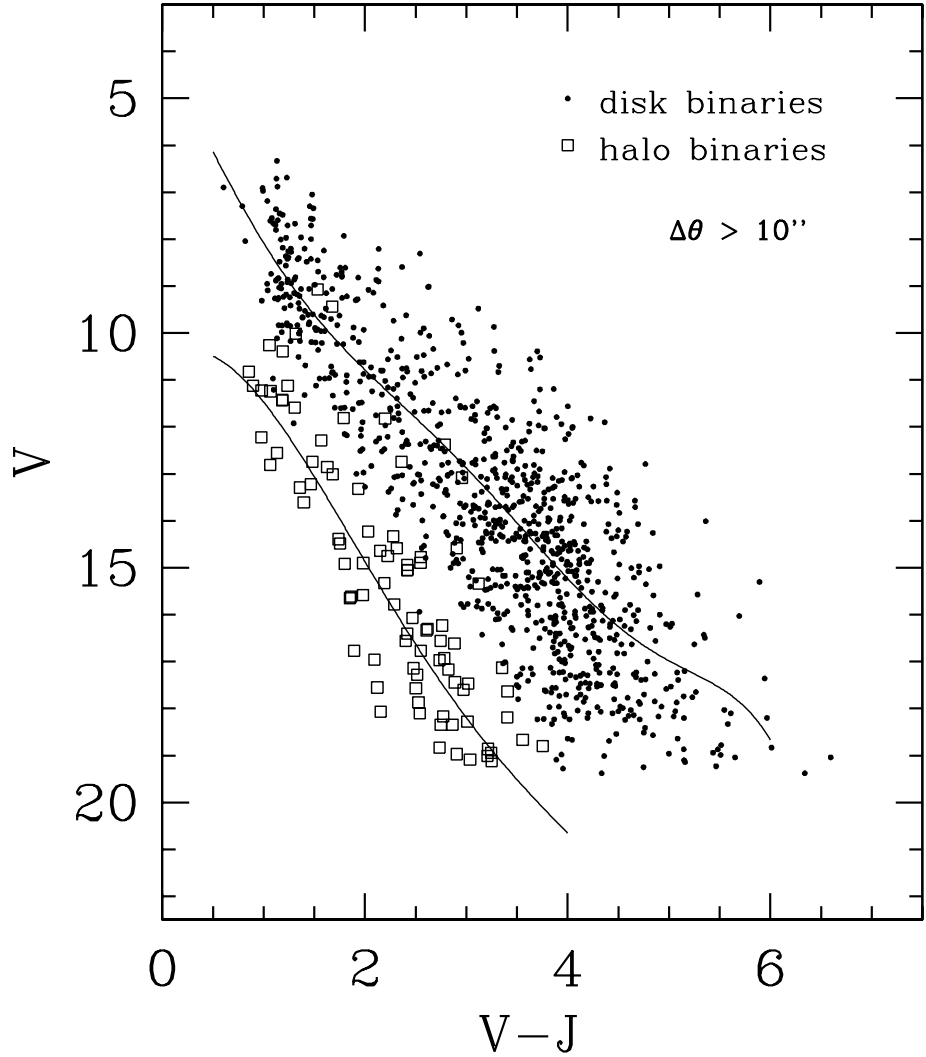}
\vspace{-2.0cm}
\caption{\label{fig:seven} 
Color-magnitude diagram of the disk and
halo samples of wide ($\Delta\theta > 10\arcsec$) binaries that
satisfactorily passed the selection and classification procedures.
The solid lines are the color-magnitude relations obtained from the
parallax samples discussed in \S 4.2, and placed at distances of 60
and 240 pc, for the disk and halo binaries respectively.
}\end{figure}

\begin{figure}
\vspace*{0cm}
\hspace*{-17cm}
\plotfiddle{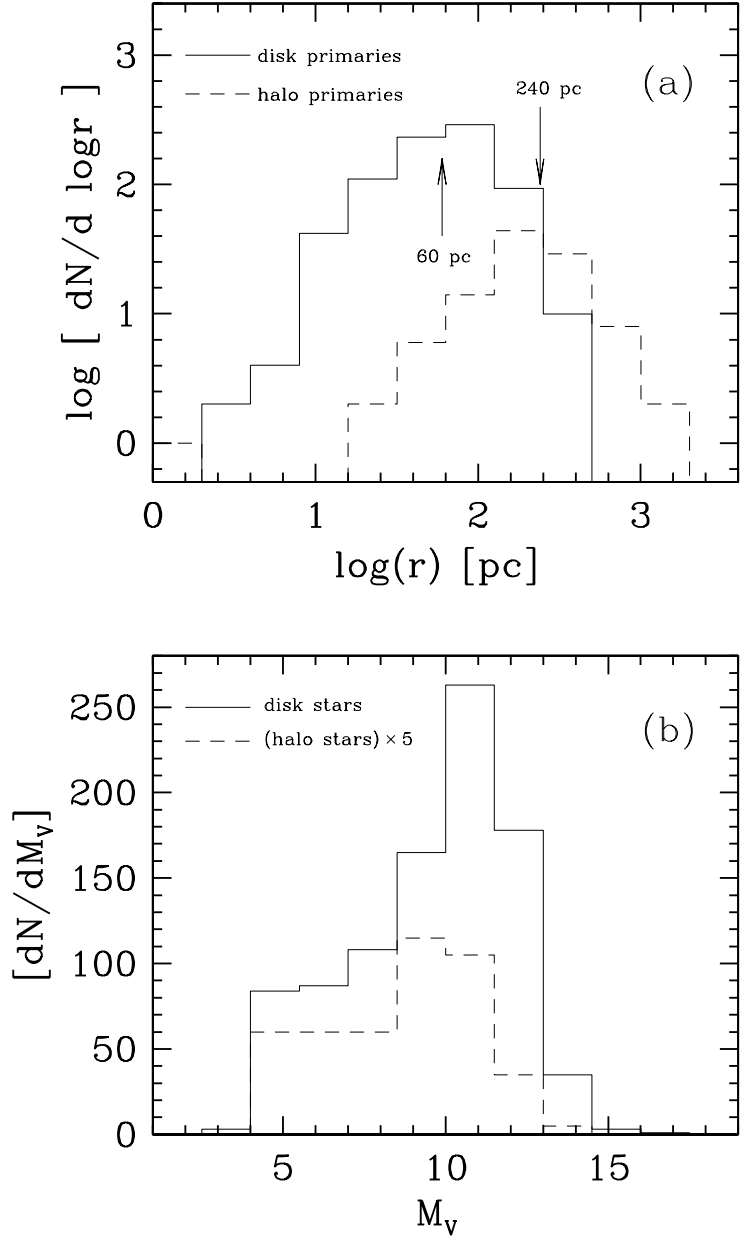}{1cm}{0}{0.7}{0.7}{0}{0}
\vspace{-1.0cm}
\caption{\label{fig:eigth} (a) Distance distribution of all the disk
and halo primaries, as obtained with the color-magnitude relations
found in \S 4.2.  Both close and wide subsamples are included.  From
these distributions, average distances of 60 and 240 pc are adopted
for the disk and halo samples, respectively.  (b) $V$-band luminosity
functions for stars in disk and halo wide ($\Delta\theta > 10\arcsec$)
binaries, including primaries and secondaries, as obtained from the
color-magnitude relations of \S 4.2.  The locations of the peaks of
the disk and halo luminosity functions are consistent with those found
for solar neighborhood M dwarfs and SDs, respectively.  }\end{figure}

\begin{figure}
\vspace{-3.0cm}
\plotone{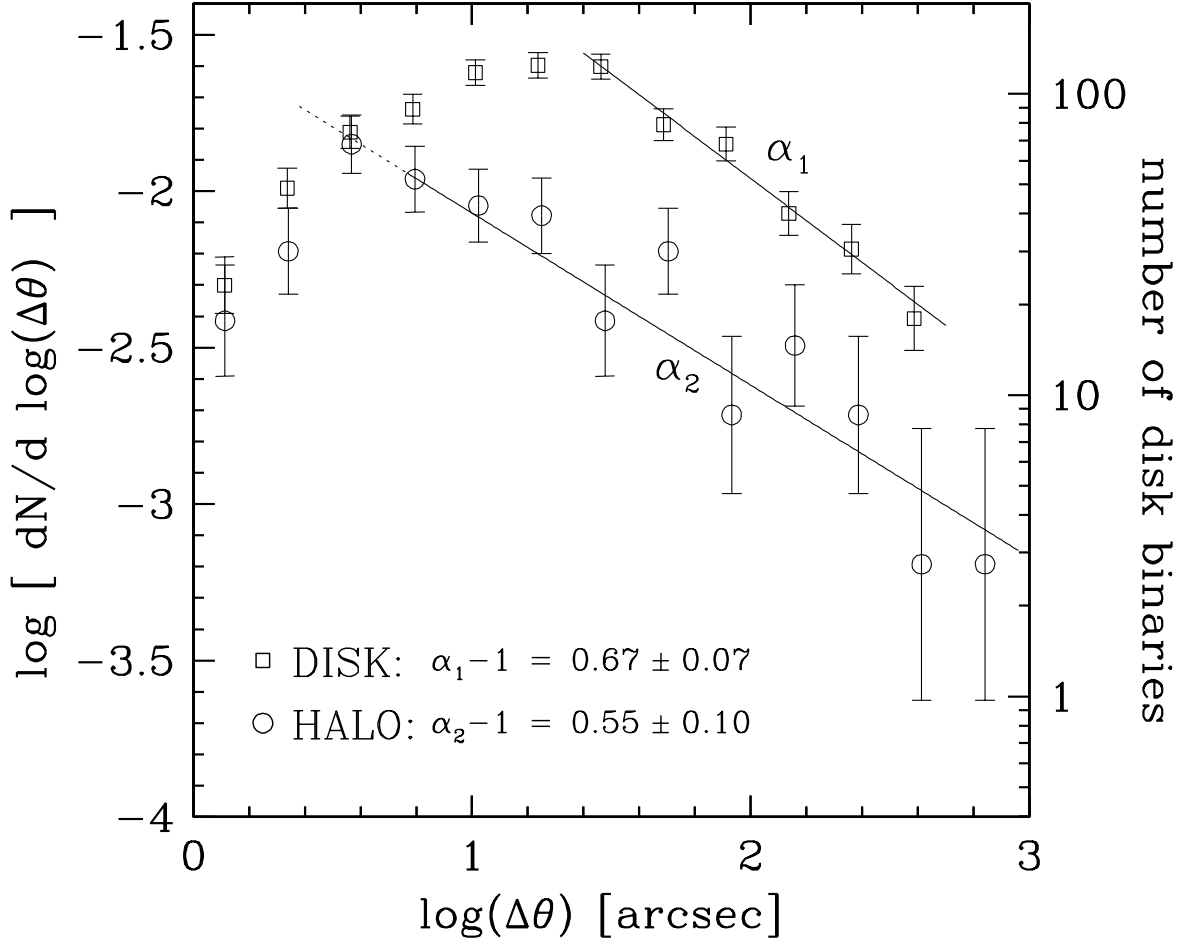}
\vspace{-2.0cm}
\caption{\label{fig:nine}
Distribution of angular separations for the final clean samples of disk
and halo binaries.  The normalizations are, respectively, with respect to 
all the disk and halo stars in the underlying rNLTT catalog.  The scale of
the right-hand vertical axis indicates the actual number of binaries in each
bin of angular separation, and refers only to the disk distribution.  The error
bars represent Poisson uncertainties.  The solid lines are the power-law
fits obtained in \S 4.3, and are limited to the range of separations used
in these fits.  The dashed extension of the halo fit at the close end 
stresses the fact that the same power law still holds to separations even
smaller than the those used in the fit.  Note the flattening of the disk
distribution between $10\arcsec-25\arcsec$, which has high statistical
significance and occurs well outside the region of image blending (see 
\S 4.3 and \S 5.2).  
}\end{figure}

\begin{figure}
\vspace*{0cm}
\hspace*{-16cm}
\plotfiddle{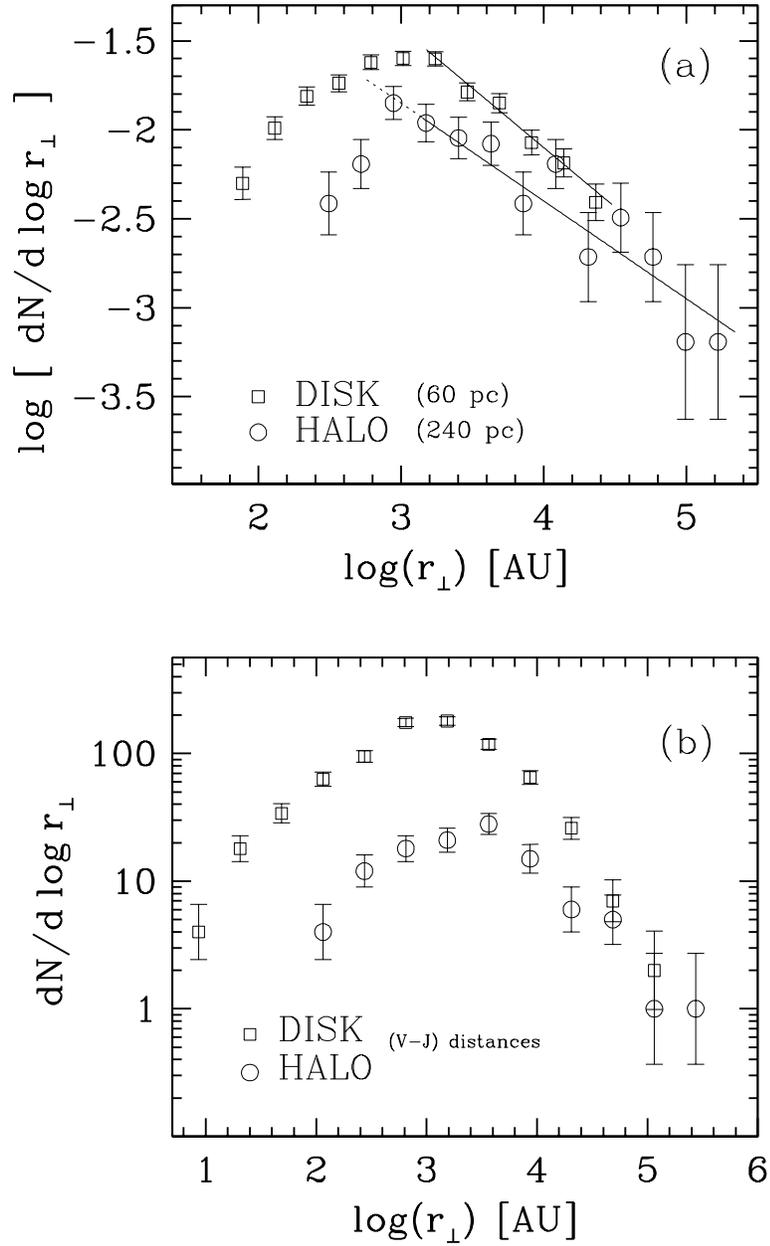}{1cm}{0}{0.7}{0.7}{0}{0}
\vspace{-1.0cm}
\caption{\label{fig:ten} Distributions of physical projected
separations for the final clean samples of disk and halo binaries, (a)
using the average distances found in Fig.~8a for the disk and halo
samples, and (b) using the individual distances to each of the
binaries as obtained from the color-magnitude relations of Figs. 6 and
7.  In (a) the normalizations relative to the entire rNLTT catalog
have been preserved, while in (b) the actual counts are shown.
}\end{figure}

\begin{figure}
\vspace*{0cm}
\hspace*{-18cm}
\plotfiddle{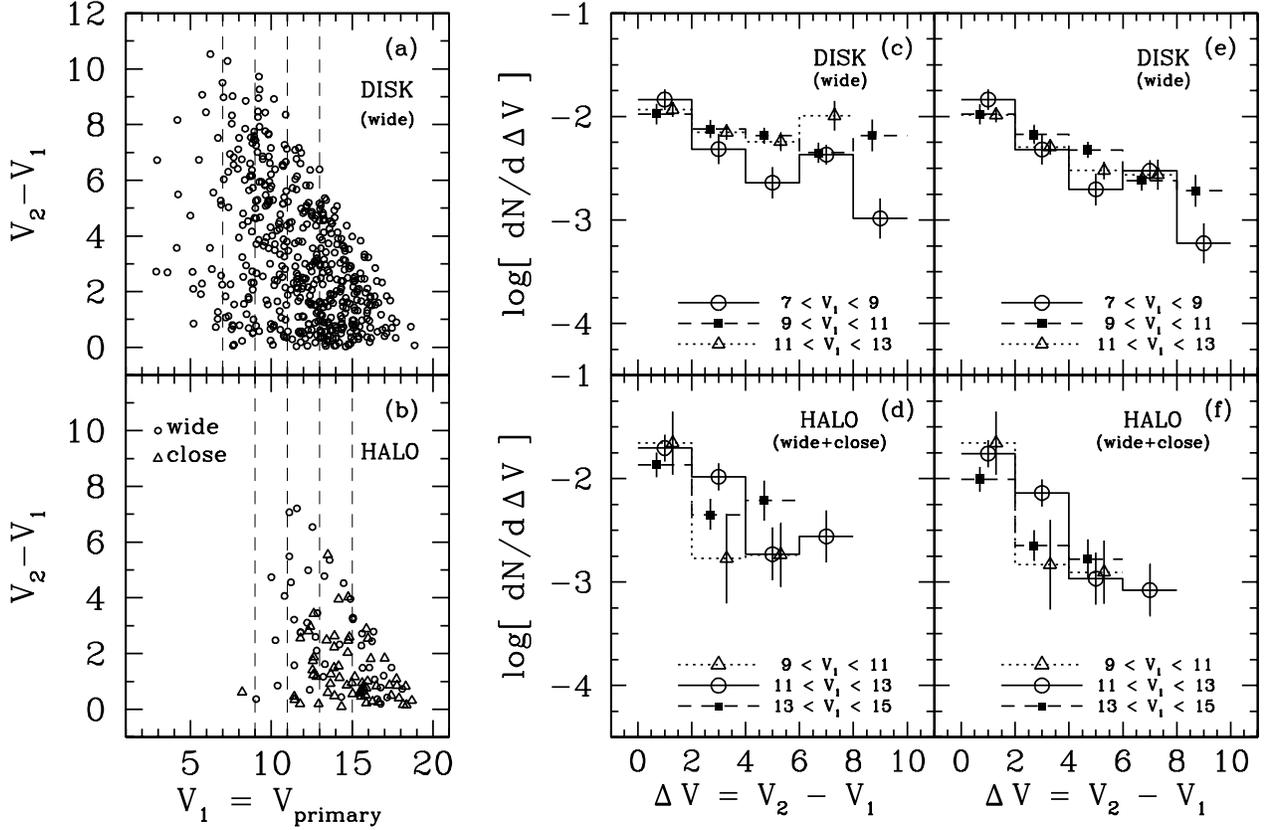}{1cm}{-90}{0.65}{0.65}{0}{0}
\vspace{-2.5cm}
\caption{\label{fig:eleven} Distributions of luminosity ratios.  (a,b)
Magnitude difference between primaries and secondaries of disk and
halo binaries as a function of the $V$-band magnitudes of the
primaries.  Only wide ($\Delta\theta > 10\arcsec$) binaries are included in
the disk diagram, while binaries at all separations are included in
the halo diagram to increase statistics.  Note the magnitude limit of
the rNLTT catalog as evidenced by the sharp upper envelope limiting
the location of the points in these diagrams.  The dashed vertical
lines indicate the regions chosen to measure the distributions shown
in the next panels.  (c,d) Distributions of luminosity ratios for the
three different ranges of primary's magnitude selected in (a) and (b),
normalized with respect to the entire rNLTT catalog (see \S4.4 for
details) with no correction for catalog incompleteness.  (e,f) Same as
(c,d) but normalizing with respect to the rNLTT catalog as corrected
for incompleteness (see \S 4.4).  All the disk and halo distributions
in panels (c) through (f) show a preference for binaries with
equal-luminosity components.  }\end{figure}

\begin{figure}
\vspace*{0cm}
\hspace*{-18cm}
\plotfiddle{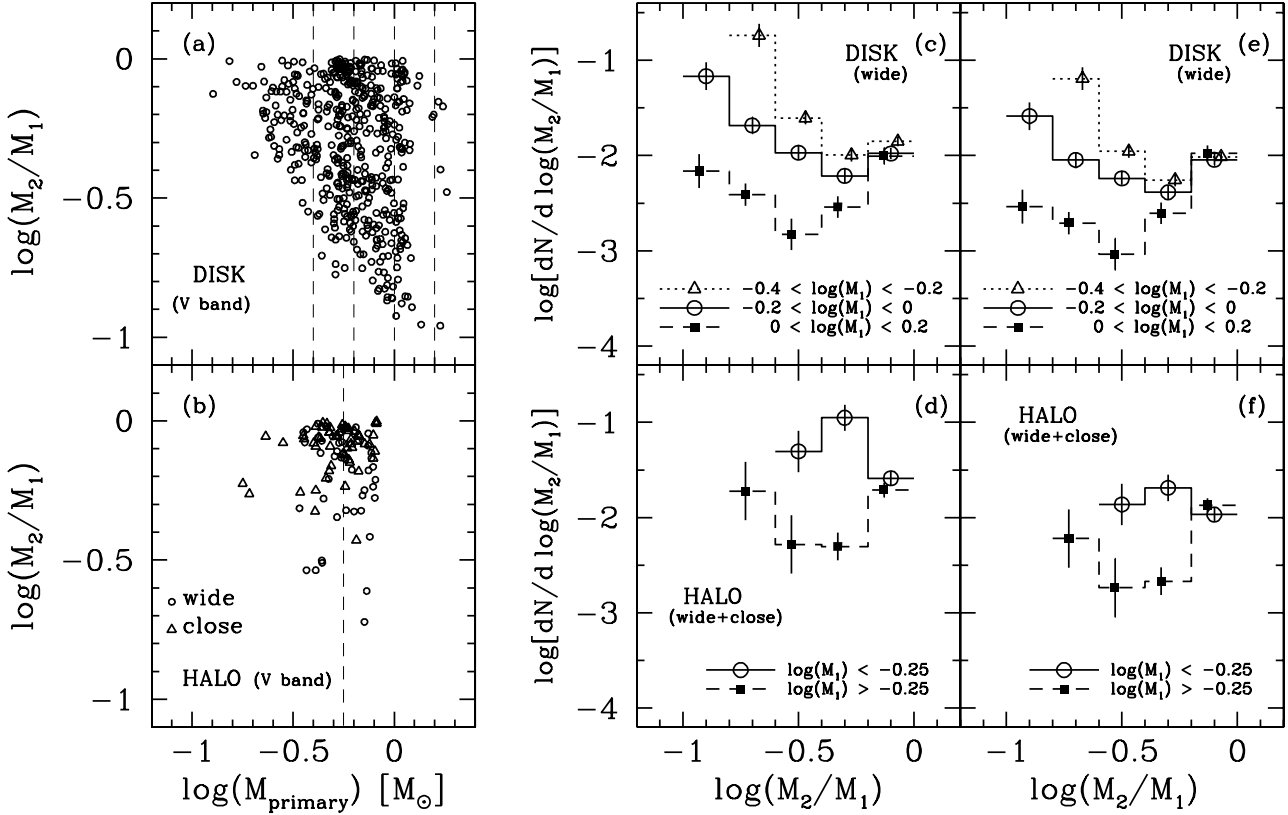}{1cm}{-90}{0.65}{0.65}{0}{0}
\vspace{-2.5cm}
\caption{\label{fig:twelve} Distributions of mass ratios.  (a,b) Ratio
of the mass of secondaries to that of their primaries as function of
primary's mass.  Only wide ($\Delta\theta > 10\arcsec$) binaries are
included in the disk diagram, while binaries at all separations are
included in the halo diagram.  The dashed vertical lines indicate the
regions chosen to measure the distributions shown in the next panels.
(c,d) Distributions of mass ratios for the ranges of primary mass
selected in (a) and (b), normalized with respect to the entire rNLTT
catalog (see \S4.4 for details) with no correction for catalog
incompleteness.  (e,f) Same as (c,d) but normalizing with respect to
the rNLTT catalog as corrected for incompleteness.  The data seem to
indicate a preference for binaries with equal-mass components, but
possible selection effects complicate the interpretation.  See \S 4.4
and \S 5.2 for an extended discussion.  }\end{figure}

\begin{figure}
\vspace*{0cm}
\hspace*{-18cm}
\plotfiddle{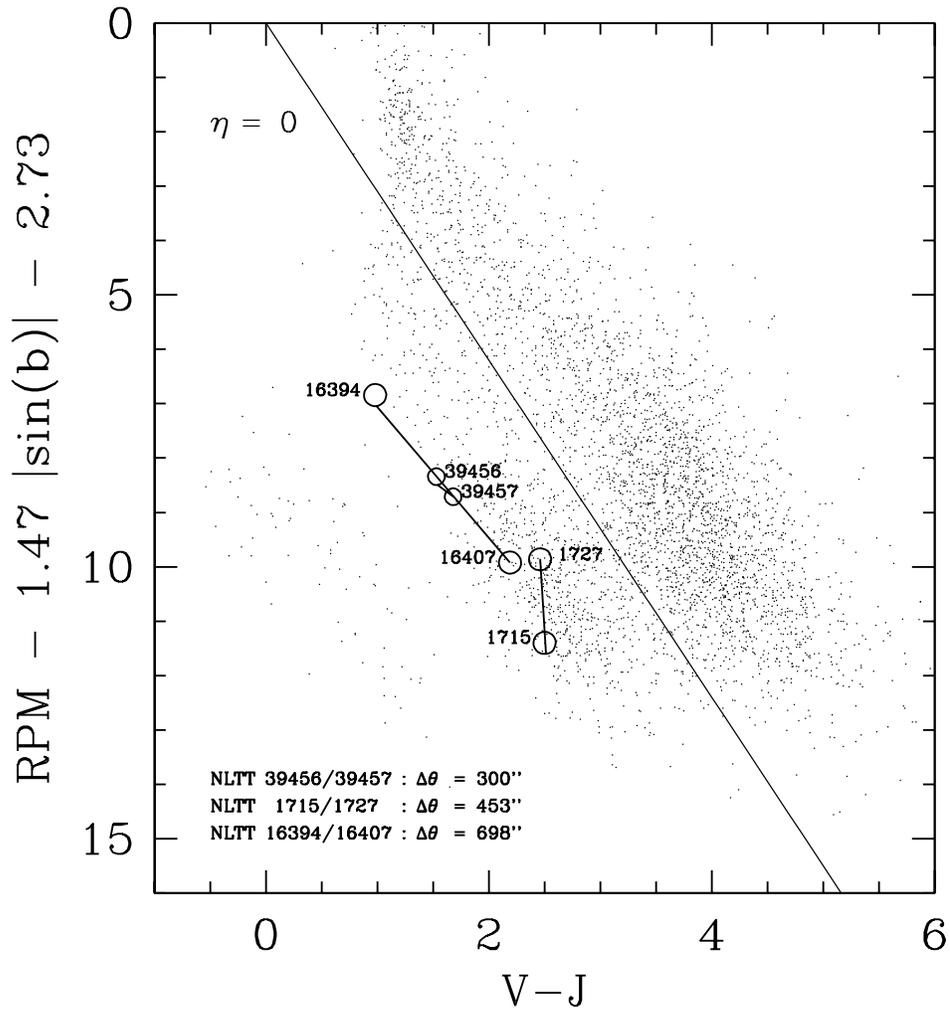}{1cm}{0}{0.7}{0.7}{0}{0}
\vspace{-2.0cm}
\caption{\label{fig:thirteen} Reduced proper-motion diagram of the
three widest binaries in the halo sample.  The CPM pairs NLTT
39456/39457 and NLTT 16394/16407 have strong evidence of being real
bound systems, while the pair NLTT 1715/1727 should be taken with
caution (see \S 5.1).  }\end{figure}

\begin{figure}
\vspace{-2.0cm}
\plotone{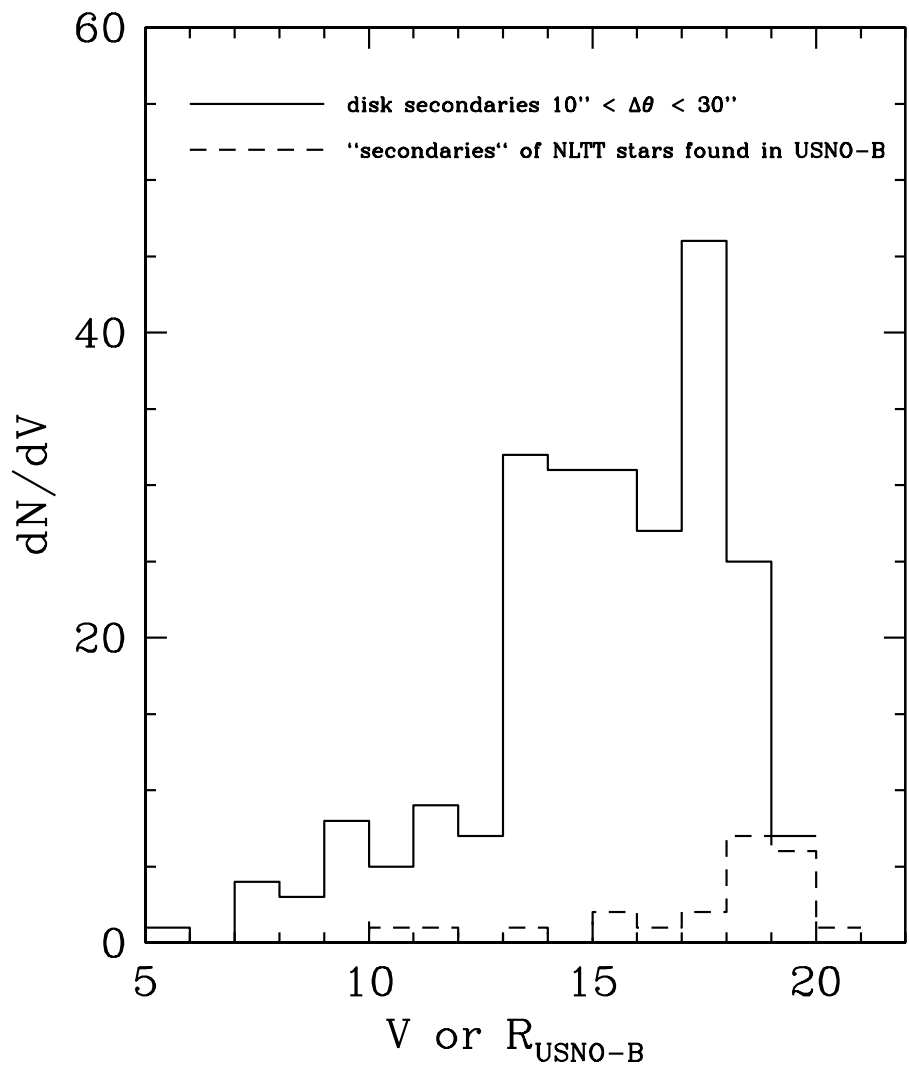}
\vspace{-2.0cm}
\caption{\label{fig:fourteen}
Search in USNO-B for non-NLTT CPM companions with separations 
$10\arcsec < \Delta\theta < 30\arcsec$
from NLTT stars.  Shown here is the distribution of USNO-B $R$ magnitudes
of the stars found (dashed histogram), in comparison to the distribution of
$V$-band magnitudes of all the secondaries of disk binaries in our sample 
(solid histogram).  The majority of the new CPM companions are very faint, 
consistent with the type of secondaries expected to be missing due to the
NLTT catalog magnitude limit. 
}\end{figure}

\begin{figure}
\vspace*{0cm}
\hspace*{-18cm}
\plotfiddle{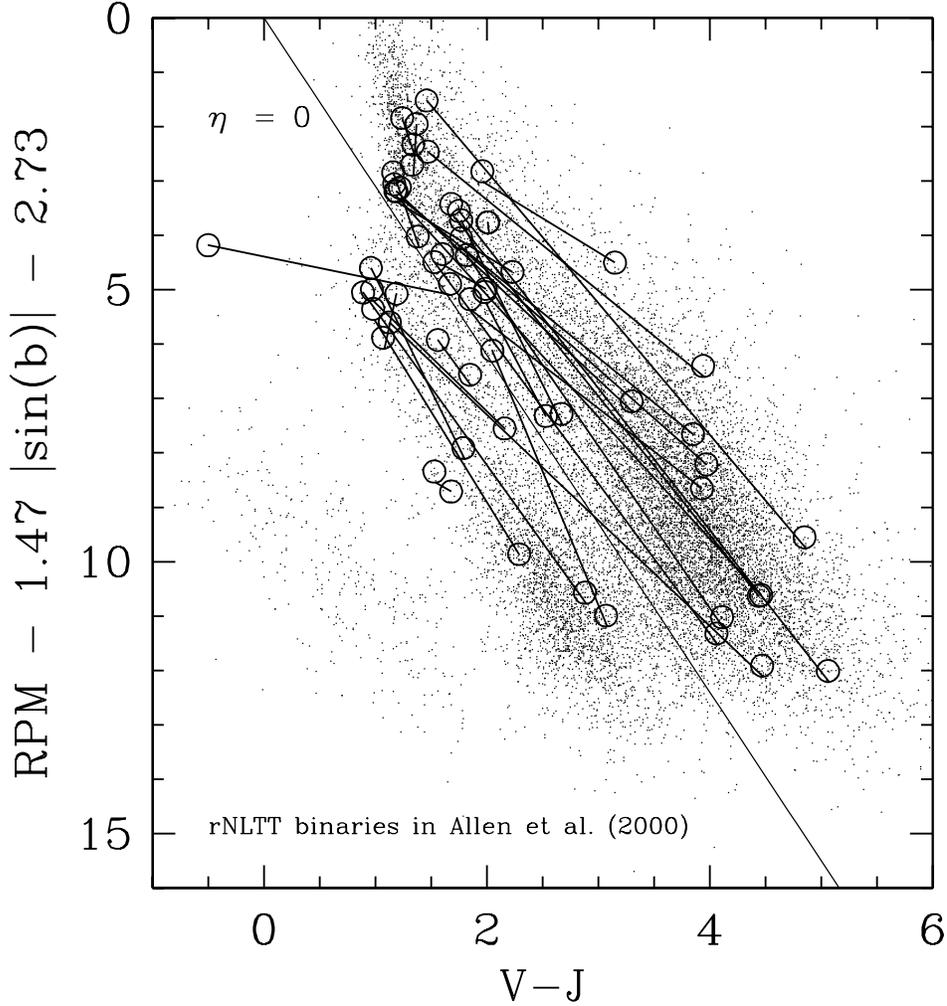}{1cm}{0}{0.7}{0.7}{0}{0}
\vspace{-2.0cm}
\caption{\label{fig:fifteen} Reduced proper-motion diagram of all the
binaries common to the work of \citet{allen} and the present one.
Comparison with the RPM diagram of Fig. 3 shows that almost all of the
binaries in \citet{allen} have G-type primaries.  Note the two
unphysical disk-halo pairs, with the lines connecting their components
crossing the disk/halo boundary ($\eta=0$).  Another pair in the upper
part of the diagram, showing a component with $V-J \sim -0.5$, is a
close binary ($\Delta\theta = 5\arcsec$) with blended USNO-A
photometry, and for this reason is classified following the Luyten
photometry (see \S 3.1).  Six of the 37 binaries in common cannot be
plotted here because one or both components do not have $J$-band
measurements available.  }\end{figure}

\begin{figure}
\vspace*{0cm}
\hspace*{-17cm}
\plotfiddle{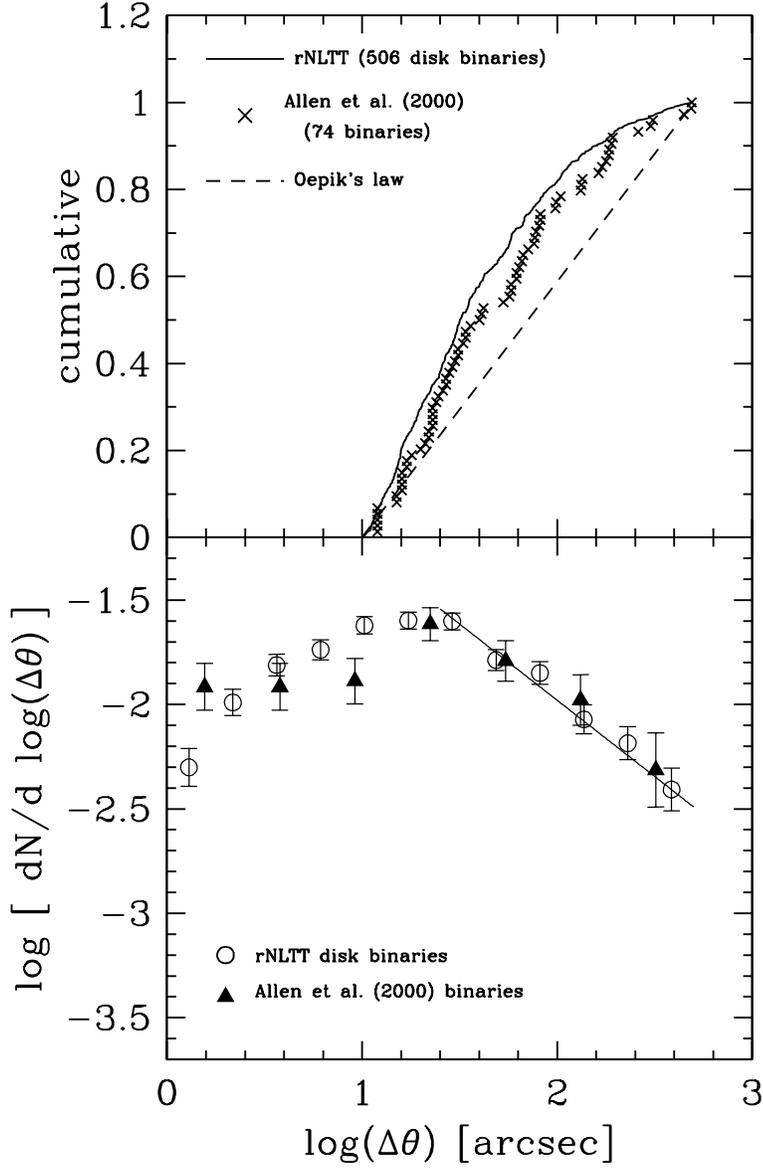}{1cm}{0}{0.7}{0.7}{0}{0}
\vspace{-1.0cm}
\caption{\label{fig:sixteen} Distributions of angular separations for
the \citet{allen} sample of binaries in comparison with that of the
disk binaries of the present work.  The upper panel shows the
respective cumulative distributions for all binaries wider than
$10\arcsec$: both samples follow similar distributions, but very
different than Oepik's law, which is represented as the dashed line.
The lower panel shows the distributions of the complete samples in
differential log-log form, where it can be seen that they match very
well.  The normalization in the lower panel is the same as in Fig. 9
for the rNLTT binaries, while for the binaries in \citet{allen} a
vertical offset is applied to emphasize the agreement between the two
distributions.
}\end{figure}

\end{document}